\documentclass[aps,prd,twocolumn,letter]{revtex4}
\usepackage{ulem}
\usepackage{color}
\usepackage{graphicx}
\usepackage{epsfig}

\usepackage{bm}
\newcommand{\be}{\begin{equation}}
\newcommand{\ee}{\end{equation}}
\newcommand{\ba}{\begin{eqnarray}}
\newcommand{\ea}{\end{eqnarray}}

\begin{document}
\title{Heavy Flavor in Medium Momentum Evolution: Langevin vs Boltzmann} 
\author{Santosh K. Das$^{a,b}$, Francesco Scardina$^{a,b}$, Salvatore Plumari$^{a,b}$, Vincenzo Greco$^{a,b}$}

\affiliation{$^a$ Department of Physics and Astronomy, University of Catania, 
Via S. Sofia 64, 1-95125 Catania, Italy}
\affiliation{$^b$ Laboratori Nazionali del Sud, INFN-LNS, Via S. Sofia 62, I-95123 Catania, Italy}

\date{}
\begin{abstract}
The propagation of heavy quarks in the quark-gluon plasma (QGP) has been
often treated within the framework of the Langevin equation (LV), i.e. assuming 
the momentum transfer is small or the scatterings are 
sufficiently forward peaked, small screening mass $m_D$. 
We address a direct comparison between the Langevin dynamics 
and the Boltzmann collisional integral (BM) when a bulk medium is in equilibrium
at fixed temperature. 
We show that unless the cross section is quite forward peaked ($m_D\cong T $) or the
mass to temperature ratio is quite large ($M_{HQ}/T \gtrsim 8-10$) there are significant differences in the evolution
of the $p-$spectra and consequently on nuclear modification factor $R_{AA}(p_T)$. 
However for charm quark we find that very similar $R_{AA}(p_T)$ between the LV and BM can be obtained, 
but with a modified diffusion coefficient by about $\sim 15-50\%$ depending on the 
angular dependence of the cross section which regulates the momentum transfer. 
Studying also the momentum spread suffered by a single heavy quarks we see that at
temperatures $T\gtrsim \, 250\,\rm MeV$ the dynamics of the scatterings is far from being of Brownian type
for charm quarks.
In the case of bottom quarks  we essentially find no differences in the time evolution of the momentum
spectra between the LV and the BM dynamics independently of the angular dependence of the cross section, 
at least in the range of temperature relevant for ultra-relativistic heavy-ion collisions.
Finally, we have shown the possible impact of this study on $R_{AA}(p_T)$ and $v_2(p_T)$ for a realistic simulation of 
relativistic HIC. For larger $m_D$ the elliptic flow can be about $50\%$ larger for the Boltzmann dynamics 
with respect to the Langevin. This is helpful for a simultaneous reproduction of $R_{AA}(p_T)$ and $v_2(p_T)$.

\vspace{2mm}

\end{abstract}
\maketitle

\section{Introduction}
One of the  primary aims of the ongoing nuclear collisions at 
Relativistic Heavy Ion Collider (RHIC) and Large Hadron Collider (LHC) energies 
is to create a new state of matter where
the bulk properties of the matter are governed by the light 
quarks and gluons \cite{Shuryak:2004cy,Science_Muller}. To characterize this new phase of matter, 
usually referred to as the Quark Gluon Plasma (QGP), we need to probe it. 
In this context, the heavy quarks (HQs), mainly charm and bottom quarks,
play a crucial role since they do not constitute the bulk part   
of the matter due to their larger mass with respect to the temperature created in ultra-relativistic
heavy-ion collisions (uRHIC's) \cite{hfr}. 
HQs are therefore considered heavy for a two-fold reason: the first, typical of particle physics, is that
the mass $M_{HQ} \gg \Lambda_{QCD}$ which makes possible the evaluation of cross section and $p_T$
spectra within next-to-lead order (NLO) \cite{Cacciari:2005rk,Cacciari:2012ny}; the second, more inherent  to plasma
physics is that $M_{HQ} \gg T$ and therefore it may be expected to decouple from the medium.  Moreover
the thermal production in the QGP is also  expected to be negligible.
HQs are quite good probes of the QGP because they are produced in the very early stage
of the collision, as their production is associated with large momentum transfer. Therefore they witness 
the entire space-time evolution of the system  and their thermal production 
and annihilation can be ignored. Furthermore HQ thermalization time
in a perturbative QCD (pQCD) framework is estimated to be of the order of 10-15 fm/c for charm and about 25-30 
fm/c for bottom \cite{hfr,moore,rappv2,bass} for the temperature range relevant for the QGP formed at RHIC and LHC.
This means that one should not expect a full thermalization of HQ in uRHIC's, 
as the lifetime of the QGP is about 4-5 fm/c at RHIC and about 10-12 fm/c at LHC. However a quantitatively 
reliable estimate of $\tau^{HQ}_{th}$ needs a thorough comparison with the experimental
observations. Therefore, HQs offer the unique opportunity as a non-fully thermalized probe hence
carrying more information on the dynamical evolution of the bulk medium.

About a decade ago the expectations were for a perturbative interaction of HQs with the medium
due to the large mass with respect to the light quarks and also to the energy scale set by the QGP 
temperature. The predictions were a $R_{AA} \approx 0.6$ for charm quarks and $R_{AA} \approx 0.8-0.9$ 
for bottom quarks in central collisions \cite{Djordjevic:2005db,
Armesto:2005mz} at intermediate $p_T$.
Furthermore the elliptic flow $v_2=\langle cos(2\phi_p)\rangle$,
a measure of the anisotropy in the angular distribution, was predicted to be quite smaller with respect to the
light hadron ones \cite{Armesto:2005mz}. The first experimental results, hence came as a surprise showing a quite small
$R_{AA}(p_T)$ similar to that of pion and a quite large $v_2(p_T)$ of single $e^\pm$ coming from D and B
mesons decay. The last were in approximate agreement with a scenario of heavy quarks almost flowing with 
the bulk medium \cite{Batsouli:2002qf,Greco:2003vf}. 
This has of course even increased the interest for the understanding of the heavy flavor dynamics in the QGP.

The propagation of HQ in QGP has been quite often treated within the framework of the
Fokker-Planck equation ~\cite{hfr,BS,DKS,moore}. The main reason is that
their motion can be assimilated to a Brownian motion due to their perturbative interaction and large
mass that should generically lead to collisions sufficiently forward peaked and/or with small momentum transfer. 
Under such constraints it is known that the Boltzmann transport equation reduces to a Fokker-Planck dynamics
\cite{BS}, which constitutes a significant simplification of in medium dynamics. 
Such a scheme has been very widely employed  \cite{DKS,moore,rappv2,rappprl,hiranov2,allhf,Gossiaux:2012ea,alberico,bass,hees,bassnpa,he1,qun}
in order to calculate the experimentally  observed nuclear suppression factor ($R_{\mathrm AA}$) 
~\cite{stare,phenixe,phenixelat,alice} and their large elliptic flow ($v_2$)~\cite{phenixelat} for the non-photonic 
single electron spectra. 

Essentially all approaches show some difficulties to predict correctly both $R_{AA}(p_T)$ and $v_2(p_T)$
and such a trait is present not only at RHIC energy (where the only single electrons coming from
both B and D have been measured \cite{phenixelat,stare,phenixe}) but also in 
the first data coming from collisions at LHC energy \cite{alice}.
A successful prediction at RHIC for both $R_{AA}(p_T)$ and $v_2(p_T)$ was achieved by including non-perturbative 
contributions~\cite{rappprl} from the quasi-hadronic bound state with a subsequent hadronization 
by coalescence and fragmentation \cite{Greco:2003vf,Greco:2003xt}.
However the uncertainty in establishing the strength of the non-perturbative effect and the fraction of B and D
feed-down into single electrons does not enable to draw definitive conclusions. 
Furthermore also in a pQCD framework supplemented by Hard Thermal Loop (HTL) scheme several progresses has been made
to evaluate realistic Debye mass and running coupling constants~\cite{gossiauxv2,alberico} and also 
different models for the expansion
of the QGP~\cite{allhf,Gossiaux:2012ea} and three-body scattering effects~\cite{ko,Das} have been implemented to improve the 
description of the data.
Along with the Fokker-Planck approach, a description of HQ within a relativistic 
Boltzmann transport approach has been developed including collisional energy loss 
\cite{gre,Uphoff:2012gb} and collisional plus radiative energy loss~\cite{gossiauxv2}. 
Also other authors have in the past and more
recently undertaken the study of charm quarks within a Boltzmann approach \cite{Younus:2013rja,Zhang:2005ni,Molnar:2006ci,our}.

To clarify the possible differences that may come from a Fokker-Planck description with respect
to  the solution of the Boltzmann collision integral, we study in this paper 
in quite some detail the similarities and differences of the
HQ dynamics in a QGP medium.
This is a first study trying to understand if there can be some ambiguity in the data interpretation coming 
from differences in the two transport approaches currently employed to investigate
the phenomenology of open heavy flavor in ultra-relativistic HIC.
Indeed the motivation of employing a Fokker-Planck approach was initially more related to the prejudice
that the momentum transfer suffered by HQ is small for both charm and bottom quarks.
On the other hand, a suppression factor $R_{\mathrm AA}$ and an elliptic flow ($v_2$) similar for light and heavy flavors,
observed experimentally, raise the suspect that the momentum transfer
may not be really sufficiently small. Therefore we 
study the impact of the approximations involved by Fokker-Planck equation by means of a direct comparison with
the full collisional integral within the framework of Boltzmann transport equation.
In particular, we focus on studying the convergence of Boltzmann dynamics to the FP as a function
of the angular dependence of the scatterings, which determines the
average momentum transfer. 
The study is conducted comparing Boltzmann and Fokker-Planck dynamics in a box bulk medium 
at fixed temperature, which allows
a better assessment of the underlying dynamics providing a solid basis for understanding the
more complex HIC dynamics. Furthermore, we will discuss both the momentum evolution of a single quark
at fixed energy and of the global momentum spectra in terms of the nuclear suppression factor.
Finally we show the impact of our study on $R_{AA}(p_T)$ and $v_2(p_T)$ for a realistic simulation of 
relativistic HIC. 

The article is organized as follows. In the next section we will briefly discuss
 the Boltzmann transport equation and the Fokker-Planck
(Langevin) one. In section III, we discuss the cross section, the drag and the diffusion coefficients 
used to calculate the HQ momentum evolution within the two transport approaches.
Section IV is devoted for the numerical results and comparison between the results obtained 
from both  Langevin and Boltzmann approaches. The impact of the Boltzmann dynamics
for a realistic simulation of relativistic HIC is presented in section V .
Section VI contain the summary and conclusions.

\section{Transport approach for Heavy Quark Dynamics}

The Boltzmann equation for the HQ distribution function can be written in a compact form
as: 
\begin{equation}
 p^{\mu} \partial_{\mu}f_{HQ}(x,p)= {\cal C}[f_{HQ}](x,p)
\label{B_E}
\end{equation}
where ${\cal{C}}[f_{HQ}](x,p)$ is the relativistic Boltzmann-like collision integral and the phase-space
distribution function of the bulk medium appears as an integrated quantity in ${\cal{C}}[f_{HQ}]$. We are interested to
the evolution of the heavy quarks 
distribution function $f_{HQ}(x,p)$. The distribution function of the bulk medium has in general to be determined 
by another set of equations that could be the Boltzmann-Vlasov equation for quark and gluons or
the hydrodynamic equations. However in the present study the bulk medium  will be just a thermal
bath at equilibrium at some temperature T, which allows for better focusing , testing and assessing the dynamics of HQs
in a Boltzmann transport dynamics respect to a Fokker-Planck one. This is a key step before studying the more complex
case of the expanding medium in uRHIC where gradients of density and temperature are involved.

It is well known that the relativistic collision integral for two-body collisions can be written in a simplified form
\cite{hfr,BS} in the following way:
\begin{eqnarray}
{\cal C}[f_{HQ}](x,p) = \int d^3k&&[ \,\omega(p+k,k)f_{HQ}(x,p+k)   
\nonumber \\ 
&& -\omega(p,k)f_{HQ}(x,p) ]
\label{C22}
\end{eqnarray}

where $\omega(p,k)$ is the rate of collisions  per unit momentum phase space of heavy quark
changing the momentum from $p$ to $p-k$. 
The first term in the integrand of Eq.~\ref{C22} represents the gain term
through collisions and the second term represents the loss out of the infinitesimal volume element 
around the momentum $p$.
HQs interact with the medium by mean of two-body collisions regulated by the scattering
matrix  of the process $g+HQ \rightarrow g+HQ$ ($\sigma_{g+HQ \rightarrow g+HQ}$), therefore
defining the relative velocity between the two colliding particles as $v_{rel}$ the 
transition rate can be written as: 

\begin{equation}
\omega(p,k)=\int \frac{d^3q}{(2\pi)^3}f_g(x,p)v_{rel}\frac{d\sigma_{g+HQ \rightarrow g+HQ}}{d\Omega}
\label{om}
 \end{equation}
where $\sigma_{g+HQ \rightarrow g+HQ}$ is generally related to the
scattering matrix $|{\cal M}_{gHQ}|^2$, in this study we will employ the well-known Combridge
cross section by a screening mass in the $t$ channel propagator (see Appendix). 
The relation between the cross sections and scattering matrix is the 
standard one:
\begin{eqnarray}
&&v_{rel}\,\frac{d\sigma_{g+HQ \rightarrow g+HQ}}{d\Omega}=\frac{1}{d_c}\frac{1}{4E_pE_q}\times \nonumber\\
&&\,\, \frac{|{\cal M}_{gHQ}|^2}{16\pi^2 E_{p-k}E_{q+k}}
\delta^0 (E_p+E_q-E_{p-k}-E_{q+k})
\label{eq:cross}
 \end{eqnarray}
that we recall because the scattering matrix is the real kernel of the dynamical evolution for both
the Boltzmann approach and the Fokker-Planck one. Of course, all the calculations discussed in the following
will originate from the same scattering matrix for both cases.

The Boltzmann equation is solved numerically dividing the space into a 
three-dimensional lattice and using the test particle method to sample 
the distribution functions.
The collision integral is solved by mean of a stochastic implementation of the collision probability 
$P=v_{rel}\sigma_{g+HQ \rightarrow g+HQ}\cdot\Delta t/\Delta x$ ~\cite{greco_cascade,Ruggieri:2013bda,Greiner_cascade,Lang}.
The code has been widely tested as regard the collision rate and the evolution of non-equilibrium
initial distributions toward the Boltzmann-Juttner equilibrium distribution both as a function of cross section, 
temperature and mass of the particles, including non-elastic collisions \cite{Scardina:2012hy} .
We have considered a bulk consisting of  only gluons for simplicity. 
The extension to light quarks is straightforward, however for our purposes it is not relevant, because 
we are anyway interested in the comparison between the Boltzmann and Langevin evolution. 
From this point of view if the scatterings happen with a gluon 
or a quark is irrelevant once the angular dependence of the collisions has been fixed to be the 
same in Boltzmann and Langevin.

\subsection{Heavy quark momentum evolution in Langevin dynamics}

The non-linear integro-differential Boltzmann equation can be significantly simplified
employing the Landau approximation whose physical relevance can be associated
to the dominance of soft scatterings with small momentum transfer $|\bf k|$ with respect
to the particle momentum $\bf p$. Namely one expands $\omega(p+k,k)f(x,p+k)$ around $k$,
\begin{eqnarray}
\omega(p+k,k)f_{HQ}(x,p+k) \approx \omega(p,k)f(x,p) \nonumber \\
+k \frac{\partial}{\partial p} (\omega f) 
+\frac{1}{2}k_ik_j \frac{\partial^2}{\partial p_i \partial p_j} (\omega f)
\label{expeq_000}
\end{eqnarray}
Inserting Eq.(\ref{expeq_000}) into the Boltzmann collision integral, Eq.(\ref{C22}), 
one obtains the Fokker Planck Equation:
\begin{eqnarray}
\frac{\partial f}{\partial t}=\frac{\partial}{\partial p_i} \left[ A_i({\bf p}) f + \frac{\partial}{\partial p_j}[B_{ij}(\bf p)]\right]
\label{eq:FP}
\end{eqnarray}
by simply defining $ A_i=\int d^3k \,w({\bf p}, {\bf k}) k_i=A({\bf p})p_i$ and $B_{ij}=\int d^3k \,w({\bf p}, {\bf k}) k_i k_j$
directly related to the so called drag and diffusion coefficient.
The Fokker-Planck equation can be solved by a stochastic differential equation 
i.e the Langevin equation~\cite{hfr,moore,bass}:
\begin{eqnarray}
dx_i &=&\frac{p_i}{E}dt, \nonumber \\
dp_i &=&-A p_i dt+C_{ij}\rho_j\sqrt{dt}
\label{lv1}
\end{eqnarray}
where $dx_i$ and $dp_i$ are the coordinate and momentum changes in each time step $dt$.
$A$ is the drag force and $C_{ij}$ is the covariance matrix describing the stochastic force
in terms of independent Gaussian-normal distributed random variables $\rho_j$, 
$P(\rho)=(2\pi)^{-3/2}e^{-\rho^2/2}$. The random variable obey the 
relation $<\rho_i \rho_j>=\delta(t_i-t_j)$ and $<\rho_j>=0$.  The 
covariance matrix is directly related to the diffusion coefficient, 
\begin{eqnarray}
C_{ij}=\sqrt{2B_0}P_{ij}^{\perp}+\sqrt{2B_1}P_{ij}^{\parallel}, 
\label{cm}
\end{eqnarray}
where 
\be
P_{ij}^{\perp}= \delta_{ij}-\frac{p_i p_j}{p^2}, P_{ij}^{\parallel}=\frac{p_i p_j}{p^2}. 
\ee
are the transverse and longitudinal tensor projectors. 
Under the common assumption, $B_0=B_1=D$, 
then Eq~(\ref{cm}) reduces to $C_{ij}=\sqrt{2D(p)} \delta_{ij}$. This is exactly valid only for $p\rightarrow 0$,
but it is usually assumed also at finite $p$ in application for HQ dynamics in the 
QGP \cite{moore,rappv2,bass,rappprl,Das,hees} .
To achieve the equilibrium distribution $f_{eq}=e^{-E/T}$ with $E=\sqrt{p^2+m^2}$ as the 
ultimate distribution one needs to adjust 
the drag coefficient $A$  in accordance with the Einstein 
relation~\cite{raf,JA}
\be
A(p)=\frac{D(p)}{ET}-\frac{D^\prime(p)}{p}.
\ee
The specification of random process depend on the 
specific choice of the momentum argument of the covariance matrix. In the present 
work we are using the pre-Ito interpretation to solve Eq.(\ref{lv1}).
We have checked that if the drag $A$ and diffusion $D$ coefficients are related by the fluctuation-dissipation
theorem (FDT) the distribution function $f({\bf p})$ converges to the Boltzmann-Juttner function $e^{-E/T}$.
However when the $A$ and $D$ are directly calculated from the scattering matrix ${\cal M}_{gHQ}$
it is not guaranteed that they fulfill the FDT. In fact the Fokker-Planck equation is just a projection
of the effect of scatterings into first (drag) and second (diffusion) moments and it cannot be 
guaranteed that the dynamics implied by the scattering processes can be fully encased into a momentum
shift plus a Gaussian fluctuations around the average momentum.
However we have checked that generally for all the cases considered the violation of the FDT is marginal
at least for momenta $p\gtrsim1.5 \rm GeV$ that is the region of interest in our following discussion.

\section{Scattering Matrix, cross section and Drag-Diffusion coefficients}

\begin{figure}[ht]
\begin{center}
\includegraphics[width=17pc,clip=true]{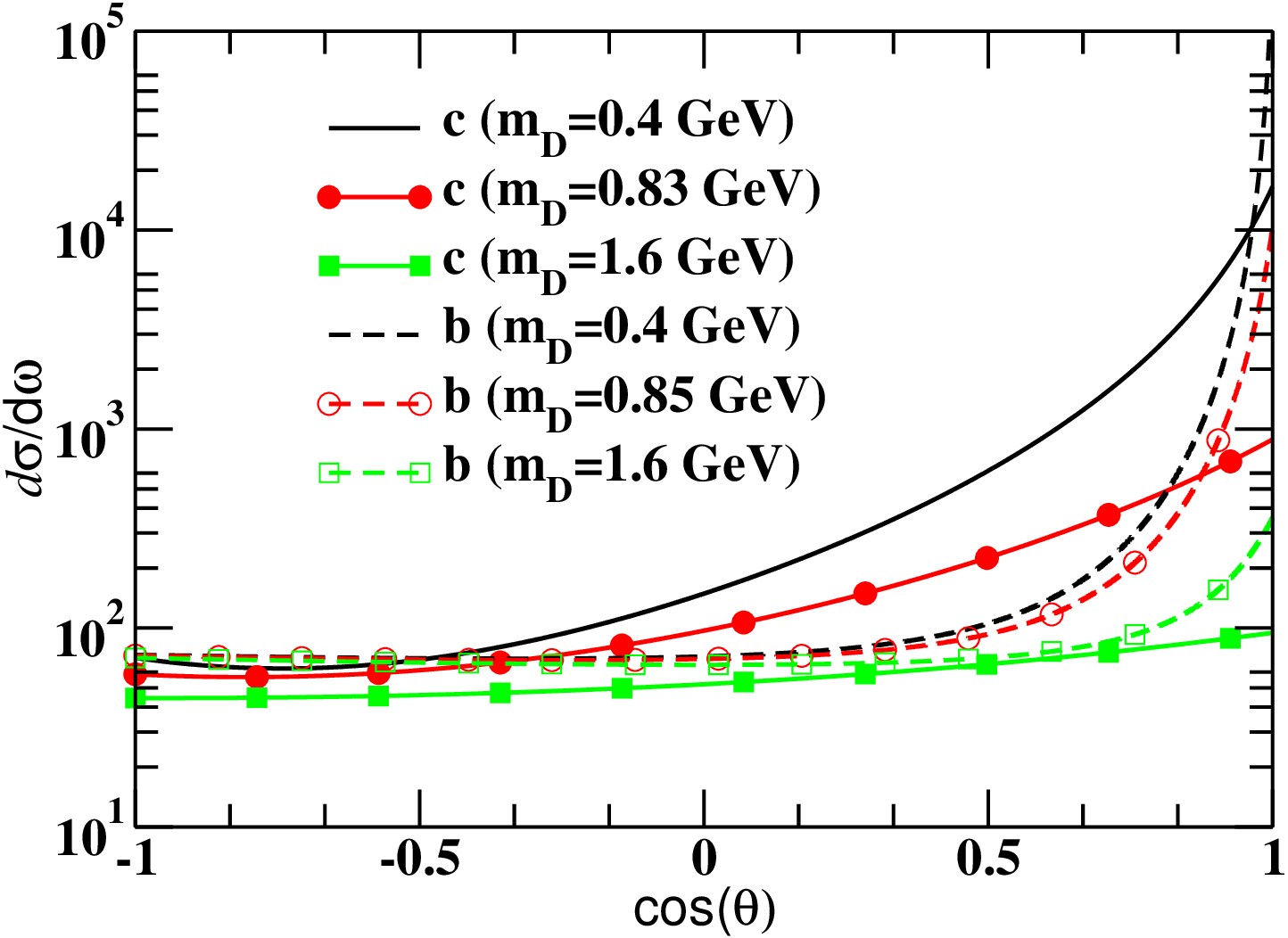}\hspace{2pc}
\caption{ Angular dependence of the cross section for different values of $m_D$  
for charm quarks (solid lines) and for bottom quarks (dashed lines).}
\label{fig1}
\end{center}
\end{figure}

The elastic collisions of heavy quarks with the gluons in the bulk has been 
considered within the framework of pQCD. 
The expression of the scattering matrix ${\cal M}_{gHQ}$ is the well known Combridge matrix
that includes $s,t,u$ channel and their interferences terms,
augmented with a screening mass $m_D=g(T) T$ inspired by the HTL scheme
as detailed in the Appendix. We have taken a charm quark mass $M_c=1.3\,\rm GeV$ and
a bottom quark mass $M_b=4.2\,\rm GeV$.
Our purpose is to perform the comparison between the Langevin and Boltzmann transport equations
for different momentum transfer scenarios that can be directly related to the angular distribution 
of scattering matrix or cross section. This can be achieved by using three different values 
of the Debye screening masses ($m_D$) needed to shield the divergence associated with the t-channel 
of the scattering matrix. 
As well known a small screening mass corresponds to forward peaked differential cross section,
as we show in Fig.\ref{fig1} by solid lines for charm quarks and dashed lines for bottom quarks.
We have chosen three different values for $m_D$, one is 0.83 GeV that corresponds to $m_D=\sqrt{4\pi\alpha_s}\, T$
with $\alpha_s=0.35$ at $T=400\,MeV$. The last is the main temperature we will consider for our study.
The other two values correspond to a reduction factor of two ($m_D=0.4\,\rm GeV$) 
and an increase of a factor of two  ($m_D=1.6\,\rm GeV$). 
We can see in Fig.\ref{fig1} that
$m_D=0.4$ GeV corresponds to a situation where the scattering is quite forward peaked and  
$m_D=1.6$ GeV instead corresponds 
to a situation where the scatterings are nearly isotropic, see Fig. \ref{fig1}. 
We consider these three cases for $m_D$ just as an effective way to roughly resemble different
modelings as the one based on very forward peaked scatterings (small $m_D$) \cite{gossiauxv2,Meistrenko:2012ju,Uphoff:2012gb}, those
more close to an HTL approach (with $m_D= gT$) \cite{Das,alberico,bassnpa,Younus:2013rja} and finally 
with large $m_D$, the physical situation in which one can predict
the existence of resonant states that corresponds to isotropic scatterings \cite{hees,rappv2,rappprl,he1}.

\begin{figure}[ht]
\begin{center}
\includegraphics[width=17pc,clip=true]{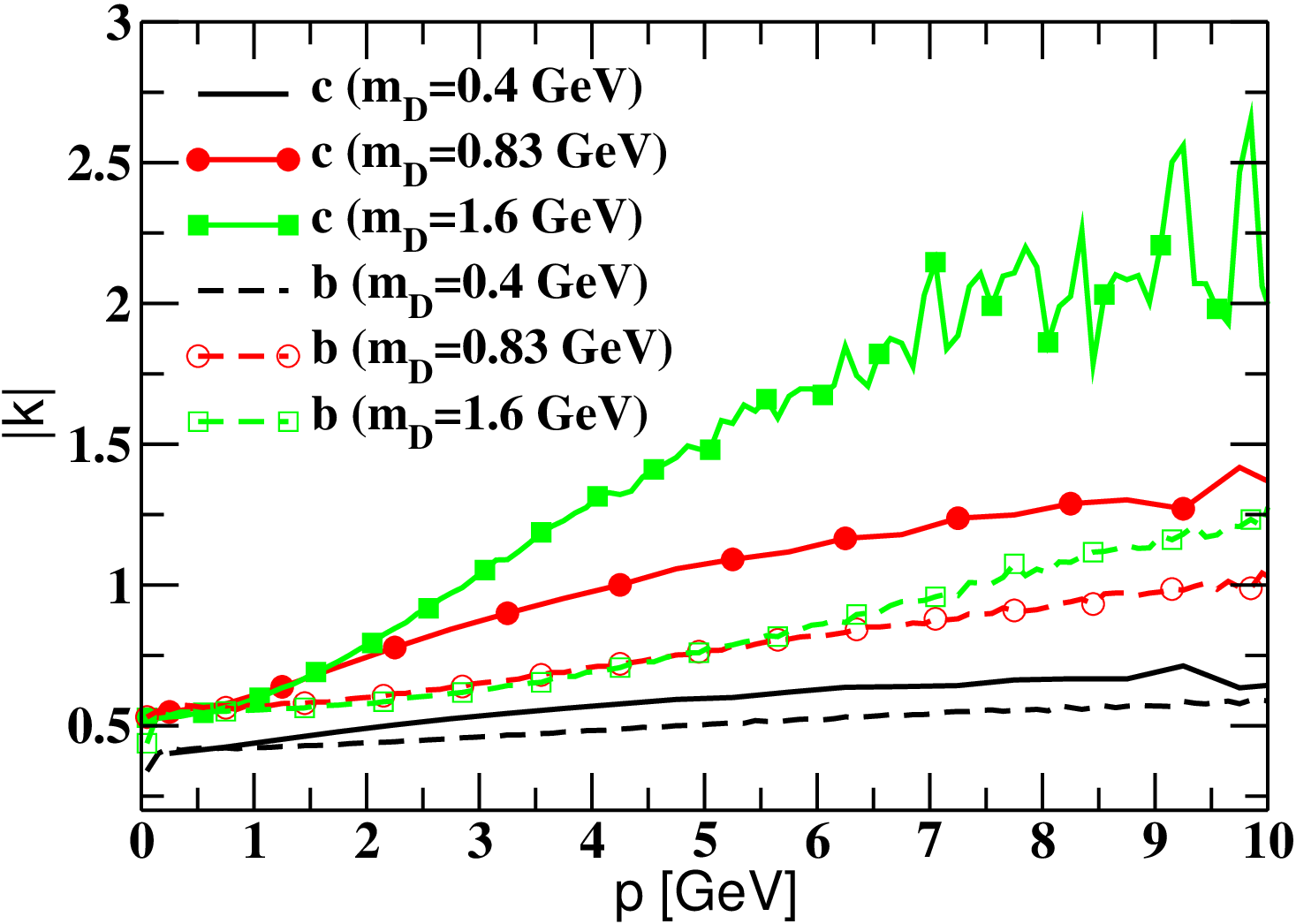}\hspace{2pc}
\caption{Variation of momentum transfer with p for different values of $m_D$ 
for charm quarks (solid lines) and for bottom quarks (dashed lines).}
\label{fig2}
\end{center}
\end{figure}

In Fig.\ref{fig2} we show the momentum transfer corresponding 
to different angular distributions or different values of Debye screening masses 
for both charm (solid lines) and bottom (dashed lines) quarks as a function of the HQ
momentum $p$ when the bulk medium is at a temperature $T=400\,\rm GeV$.
We have estimated the momentum transferred
as a function of p using the Boltzmann approach. We evaluate the total momentum 
transferred for particles in each interval $p+\Delta p$ 
and divide it by the total number of collisions in such an interval.   
For an HQ momentum $|{\bf p}|=5\,\rm GeV$ we see that one goes from a momentum
transfer $|{\bf k}|=0.5\,\rm GeV$ for the forward peaked scattering matrix corresponding to
$m_D=0.4 \,\rm GeV$ to a $|{\bf k}|=1.5\,\rm GeV$ for the nearly isotropic cross section
corresponding to $m_D=1.6 \,\rm GeV$. For bottom quarks, due to their larger mass corresponding
to $M_{b}/T\sim 10$, the change is less pronounced and
for nearly isotropic cross sections is at most about 0.8 GeV.

\begin{figure}[ht]
\begin{center}
\includegraphics[width=17pc,clip=true]{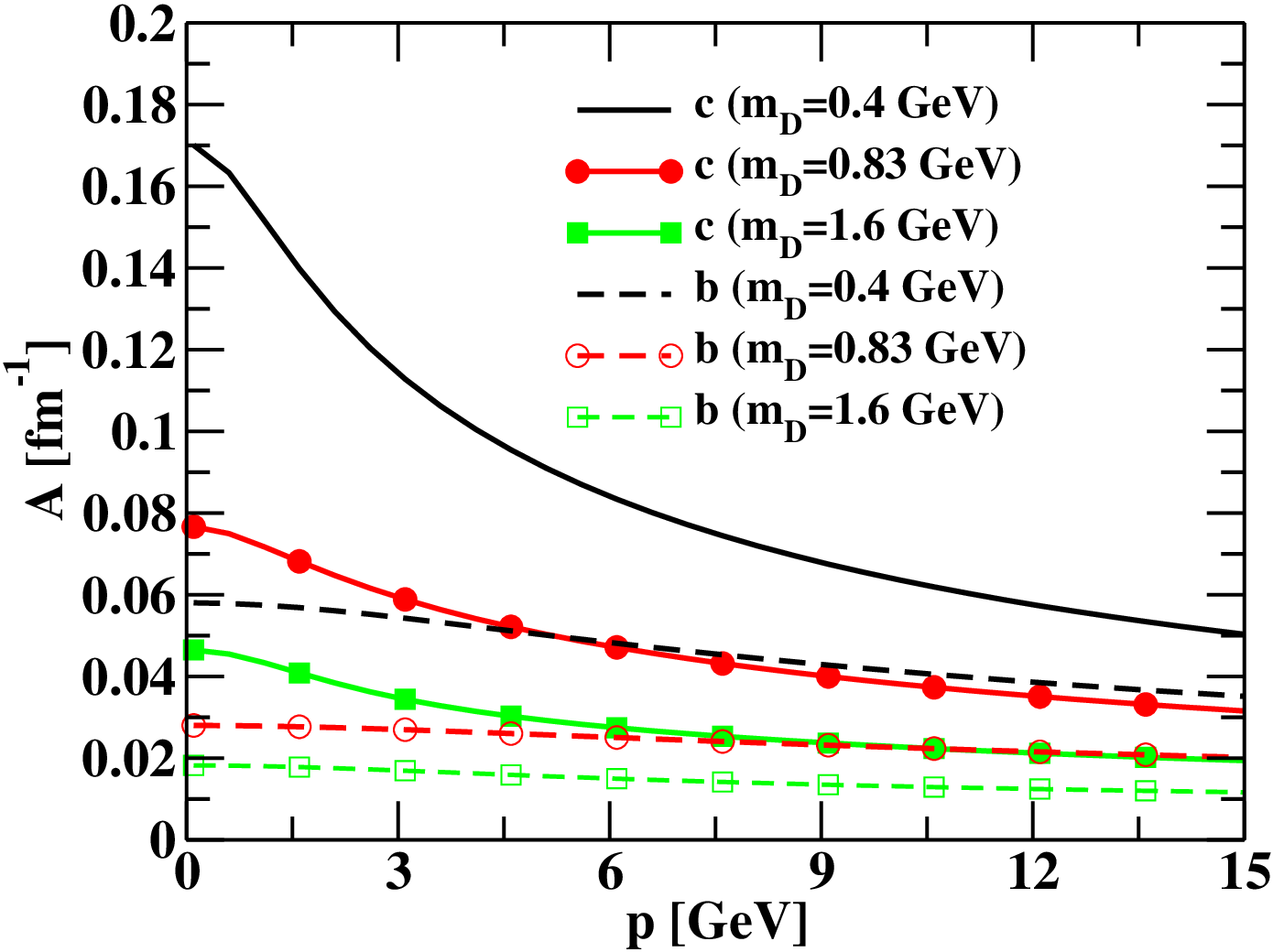}\hspace{2pc}
\caption{Variation of drag coefficients with p at T = 400 MeV for different values of $m_D$. }
\label{fig3}
\end{center}
\end{figure}

\begin{figure}[ht]
\begin{center}
\includegraphics[width=17pc,clip=true]{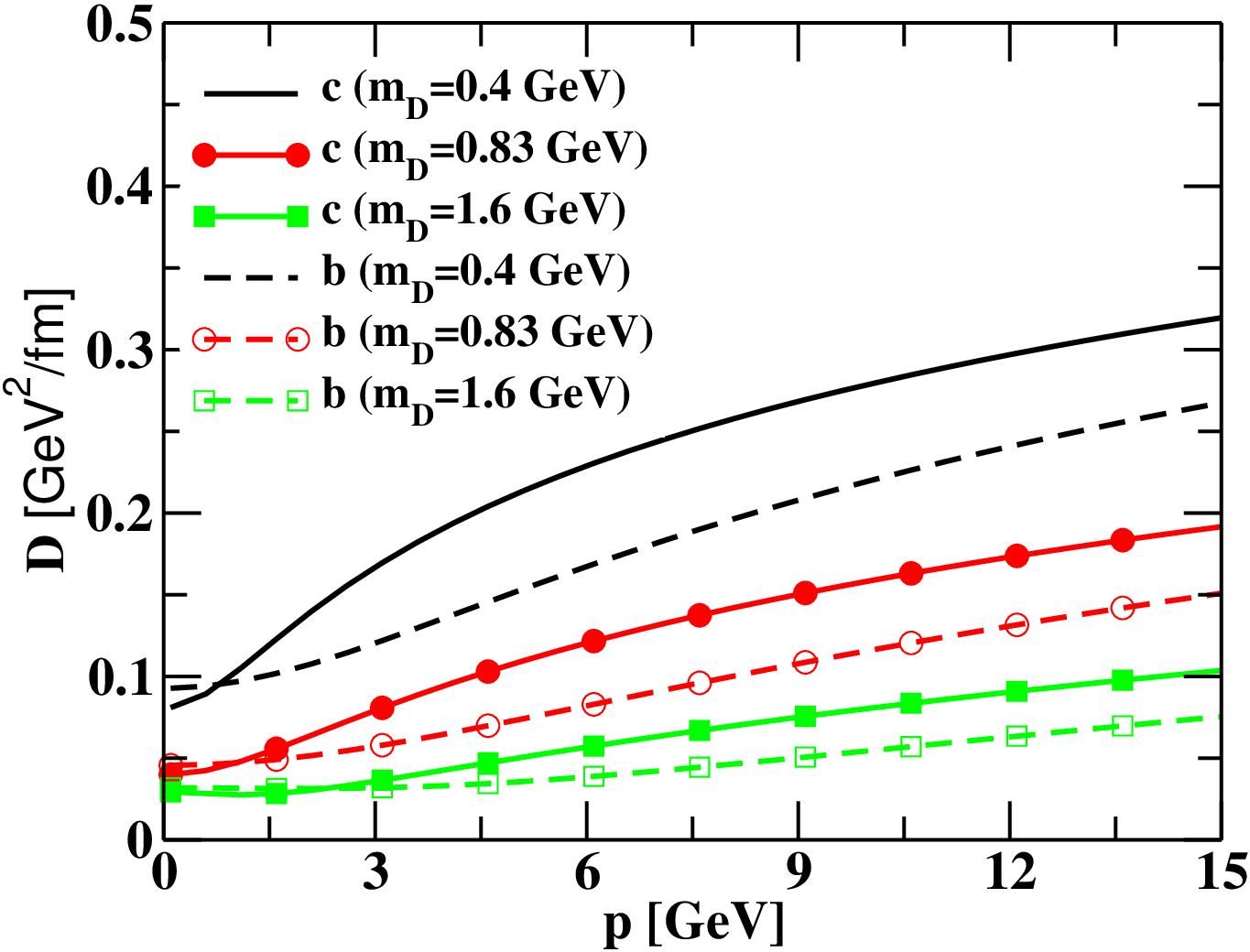}\hspace{2pc}
\caption{Variation of diffusion coefficients with p at T = 400 MeV for different values of $m_D$.}
\label{fig4}
\end{center}
\end{figure}

Starting from the same scattering matrix, ${\cal M}_{gHQ}$, we have evaluated the Drag A(p) and
Diffusion constant B(p) that enter into Langevin equation, see Eq. \ref{eq:FP}. 
The results are shown in Fig.s \ref{fig3} and \ref{fig4} for both charm (solid lines) and bottom (dashed lines) quarks
at a temperature $T = 400$ MeV for different values of  $m_D$ that give rise to different values
of drag and diffusion coefficients.  
The coupling $\alpha_s$ has been kept fixed for all the cases.
In order to have a similar $R_{AA}$ within the typical time
scale of uRHIC $\tau \approx 6 \rm\,  fm/c$, the $|{\cal M}_{gHQ}|^2$  has been multiplied by a k factor. 
This has been chosen to be
$k=2.1$ for $m_D=0.4 \rm\, GeV$, $k=4$ for $m_D=0.83\,\rm GeV$ and $k=7.2$
for $m_D=1.6\,\rm GeV$~\cite{gre}. For the same $k$ factor is of course rescaled also the cross section
used to determine the evolution within the Boltzmann equation.
Once the coefficients are rescaled by $k$, they become approximately equal at $p \sim 2-3$ GeV for the different $m_D$.
However the effect discussed does not depend on such $k$ factor that has been included only to
set similar time scales for the evolution of the spectra reaching $R_{AA}(p) \sim 0.3$
similar to that observed  in HIC at momenta
of about $4-6 \rm\, GeV$. However there is no necessity to set them exactly equal because 
we are interested only in comparing the Fokker-Planck
and Boltzmann evolution starting from the same kernel (same $m_D$) given by the scattering matrix.


\section{Numerical results: comparing the Boltzmann and Langevin evolution}

We now discuss the evolution of momentum distributions of charm and bottom quarks 
interacting with a bulk medium at $T=0.4 \,\rm GeV$ with scattering processes determined by the 
scattering matrices discussed in the previous section. The initial distribution of heavy quarks 
are taken from Ref.~\cite{Cacciari:2005rk} and given 
by $f(p,t=0)=(a+b\,p)^{-n}$ with $a=0.70\,(57.74)$, $b=0.09\,(1.00)$ and $n=15.44\,(5.04)$
for charm and bottom quarks respectively. The above function gives a reasonable description 
of D and B meson spectra in the p-p collision at highest RHIC energy. For the sake 
of comparison, we solve both the Langevin equation  and the Boltzmann
equation as described in Section II and III
in a  box with a volume $V=125 \, \rm fm^3$.   
Our purpose is to compare the time evolution starting from the same initial momentum
distribution for the both cases. The differential cross section $d\sigma/d\Omega$, main
ingredient of the Boltzmann equation,
and the drag and diffusion coefficients, key ingredient of the Langevin 
equation, both originated from the same scattering matrix. 

We have plotted the results as a ratio between Langevin to Boltzmann
at different times to quantify how much the ratio deviates from 1. We started the simulation 
at $t=0$ fm/c  which of course corresponds to a ratio of 1 as we start the simulation with the 
same initial momentum distribution 
for both Langevin and Boltzmann equations. So any deviation from 1 would reflect how much the 
Langevin differ from the Boltzmann evolution.

\begin{figure}[ht]
\begin{center}
\includegraphics[width=17pc,clip=true]{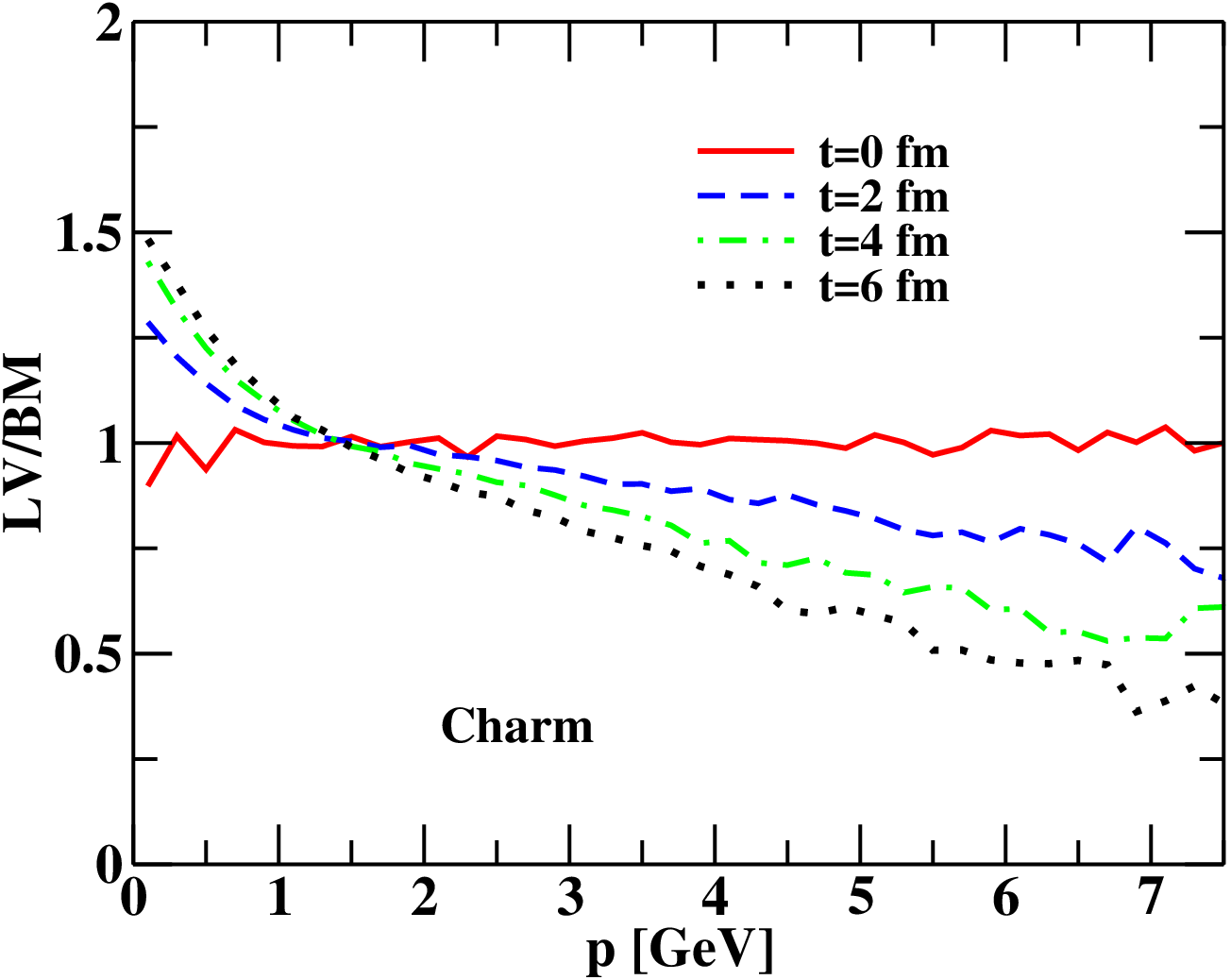}\hspace{2pc}
\caption{Ratio between the Langevin (LV) and Boltzmann (BM) spectra for charm quark 
as a function of momentum for $m_D=0.83$ GeV at different time.}
\label{fig5}
\end{center}
\end{figure}

\begin{figure}[ht]
\begin{center}
\includegraphics[width=17pc,clip=true]{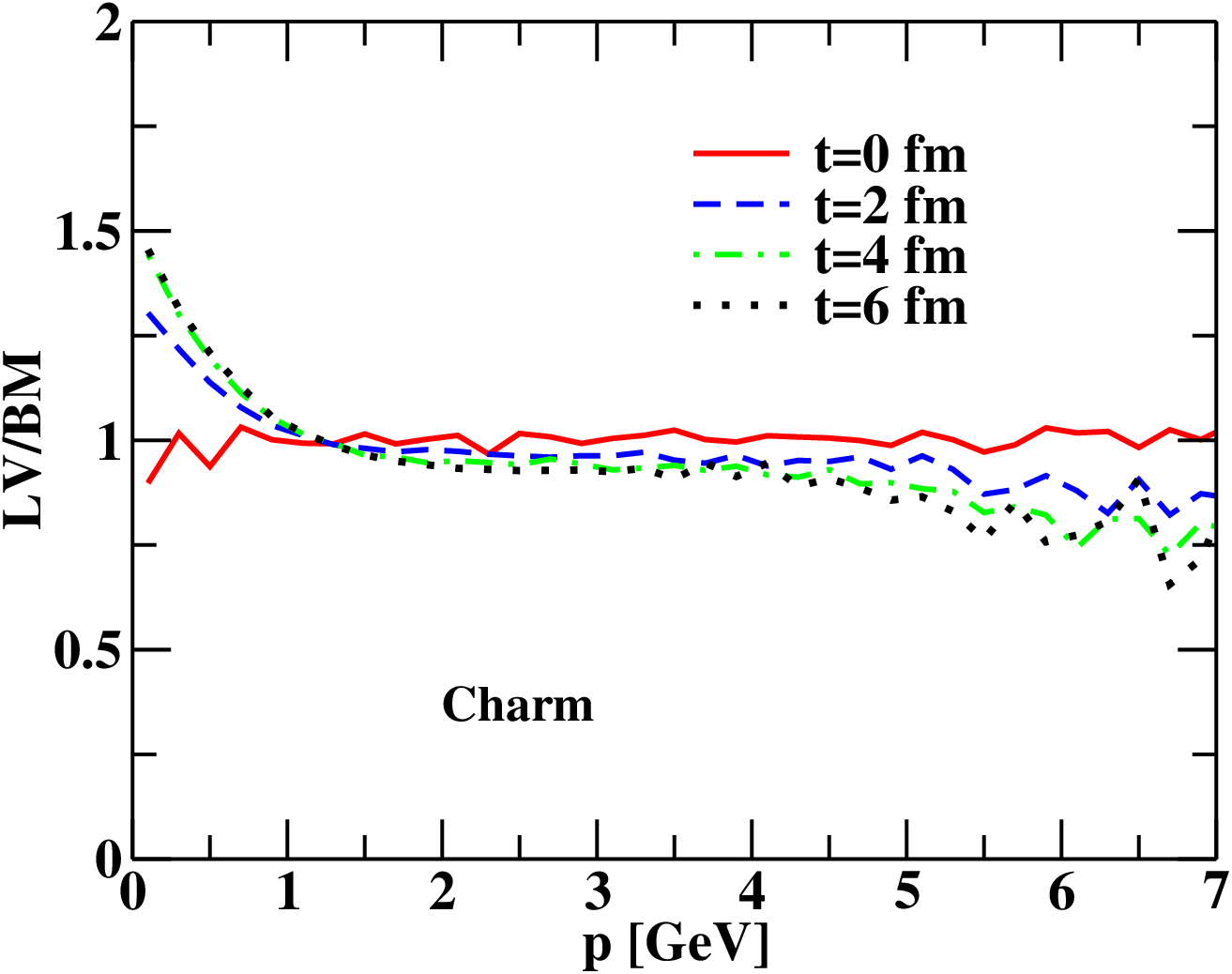}\hspace{2pc}
\caption{Ratio between the Langevin (LV) and Boltzmann (BM) spectra for charm quark 
as a function of momentum for $m_D=0.4$ GeV at different time.}
\label{fig6}
\end{center}
\end{figure}

\begin{figure}[ht]
\begin{center}
\includegraphics[width=17pc,clip=true]{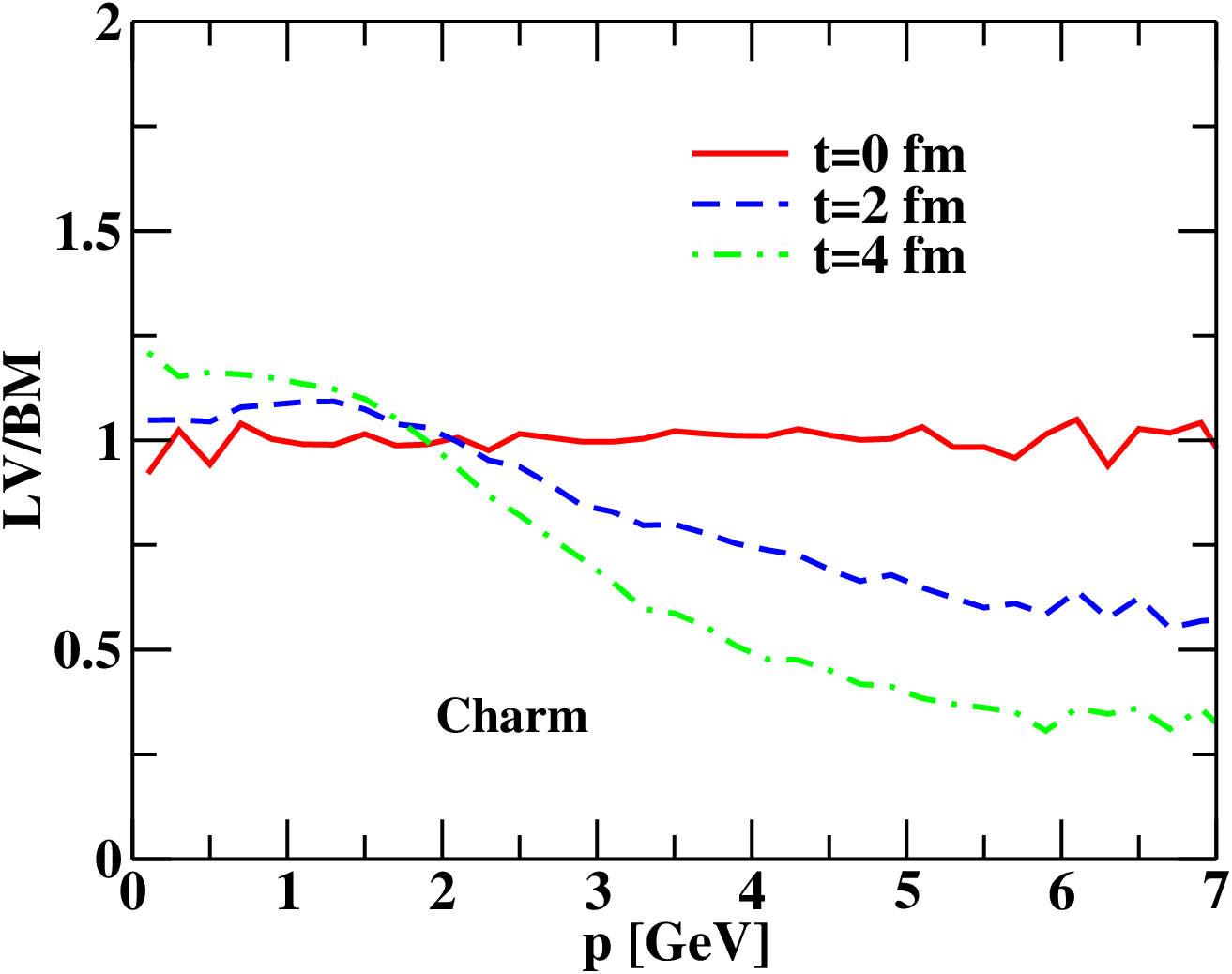}\hspace{2pc}
\caption{Ratio between the Langevin (LV) and Boltzmann (BM) spectra for charm quark 
as a function of momentum for $m_D=1.6$ GeV at different time.}
\label{fig7}
\end{center}
\end{figure}


In Fig~\ref{fig5} the ratio of Langevin to Boltzmann spectra for the charm quark with $m_D=0.83$ GeV has been displayed 
as a function of momentum at different time.    
From Fig~\ref{fig5} it is observed that for $t=4$ fm the deviation 
of Langevin from Boltzmann is around $40\%$ and for $t=6$ fm the deviation is around a $50\%$ at $p=5$ GeV,  
which suggests the Langevin approach overestimates the average energy loss considerably due to the approximation it involves.

We remind that time scales of $4-6$ fm/c can be roughly taken as those corresponding to the typical
lifetime of a QGP in uRHIC's. This is why we are displaying and discussing the results around such a time.
The same ratio is depicted in Fig~\ref{fig6} for $m_D=0.4$ GeV which is only a way to simulate a quite 
forward peaked scattering. 
The results presented 
in Fig~\ref{fig6} shown that Langevin dynamics deviates from the 
Boltzmann by about a $15\%$ at $t=6$ fm and $|{\bf p}|\approx \, 5\rm\, GeV$. The 
reduced deviation between Langevin and Boltzmann for $m_D=0.4$ GeV can be expected knowing that for
such a case where the scattering are more forward peaked and the average transfer momentum $|\bf k|$ is about 
a factor of two smaller with respect to the case  $m_D=0.83$ GeV, as shown in Fig.\ref{fig2}.
On the contrary, when we consider a larger screening mass, $m_D=1.6$ GeV to simulate a nearly isotropic scattering, 
the transferred momentum is
about a factor of three larger and we see that the ratio of Langevin to Boltzmann spectra
in Fig~\ref{fig7} at different time can lead to differences as large as a factor $70\%$ at t=4 fm/c.
It is however important to report that for this last case the results depend on the procedure chosen to determine
the Drag A(p) and Diffusion coefficients D(p). The results shown in Fig.\ref{fig7} is for the case where both coefficients 
are evaluated directly from the ${\cal M}_{gHQ}$.  
However just for this case if one calculates the diffusion coefficient from the scattering matrix
and the drag one from the constraint of the fluctuation dissipation theorem (FDT) the result are significantly
modified. In particular the LV/BM ratio will evolve quite slowly and the ratio reaches value about 
$\sim 0.6-0.7$ at t=4 fm/c. Such ambiguity in determining the drag and diffusion coefficient is much less relevant
for the case of smaller $m_D$, however it essentially means that for nearly isotropic scatterings associated 
to large momentum transfer the dynamics of the scattering cannot be really encased simply into a shift
of the average momenta with a Gaussian diffusion round the mean.  This manifests into a stronger
breaking of the FDT when both drag and diffusion are evaluated from ${\cal M}_{gHQ}$. 


\begin{figure}[ht]
\begin{center}
\includegraphics[width=17pc,clip=true]{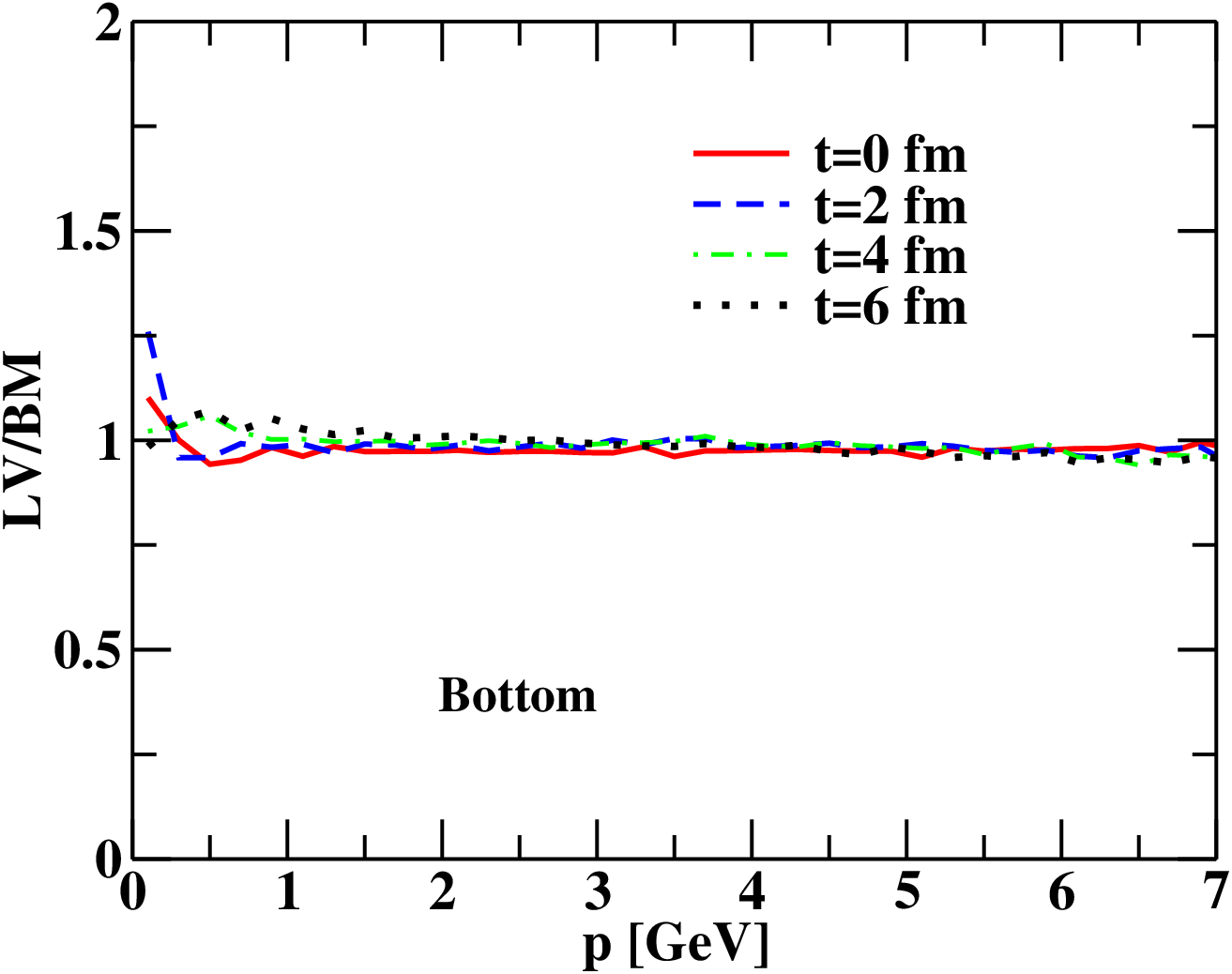}\hspace{2pc}
\caption{Ratio between the Langevin (LV) and Boltzmann (BM) spectra for bottom quark
as a function of momentum for $m_D=0.83$ GeV at different time.}
\label{fig8}
\end{center}
\end{figure}



We now move to the calculation for bottom quarks.
In Fig.~\ref{fig8} the results for bottom quark are displayed for $m_D=0.83$ GeV.
It is observed that the ratio stays practically around one for bottom quark for all the time evolution 
considered in the manuscript. 
The results for the other two values of $m_D$ are quite similar
Therefore, for bottom quark the Langevin approach is really a good approximation of the Boltzmann equation independently
of the angular dependence of the scatterings, at most a $10\%$ difference is observed for $m_D=1.6$ GeV.
On the other hand, as already mentioned due to the large bottom mass an approximation of the dynamics to a Brownian
motion appears always appropriate. We notice that this is determined by the ratio of the mass and the temperature that 
determines the average momentum of the particles colliding with heavy quarks ($\langle p\rangle \simeq 3T$). 
For the bottom quark $M_c/T\simeq 10$ 
while for charm quark $M_b/T \simeq 3$ for the temperature we are considering.

\begin{figure}[ht]
\begin{center}
\includegraphics[width=17pc,clip=true]{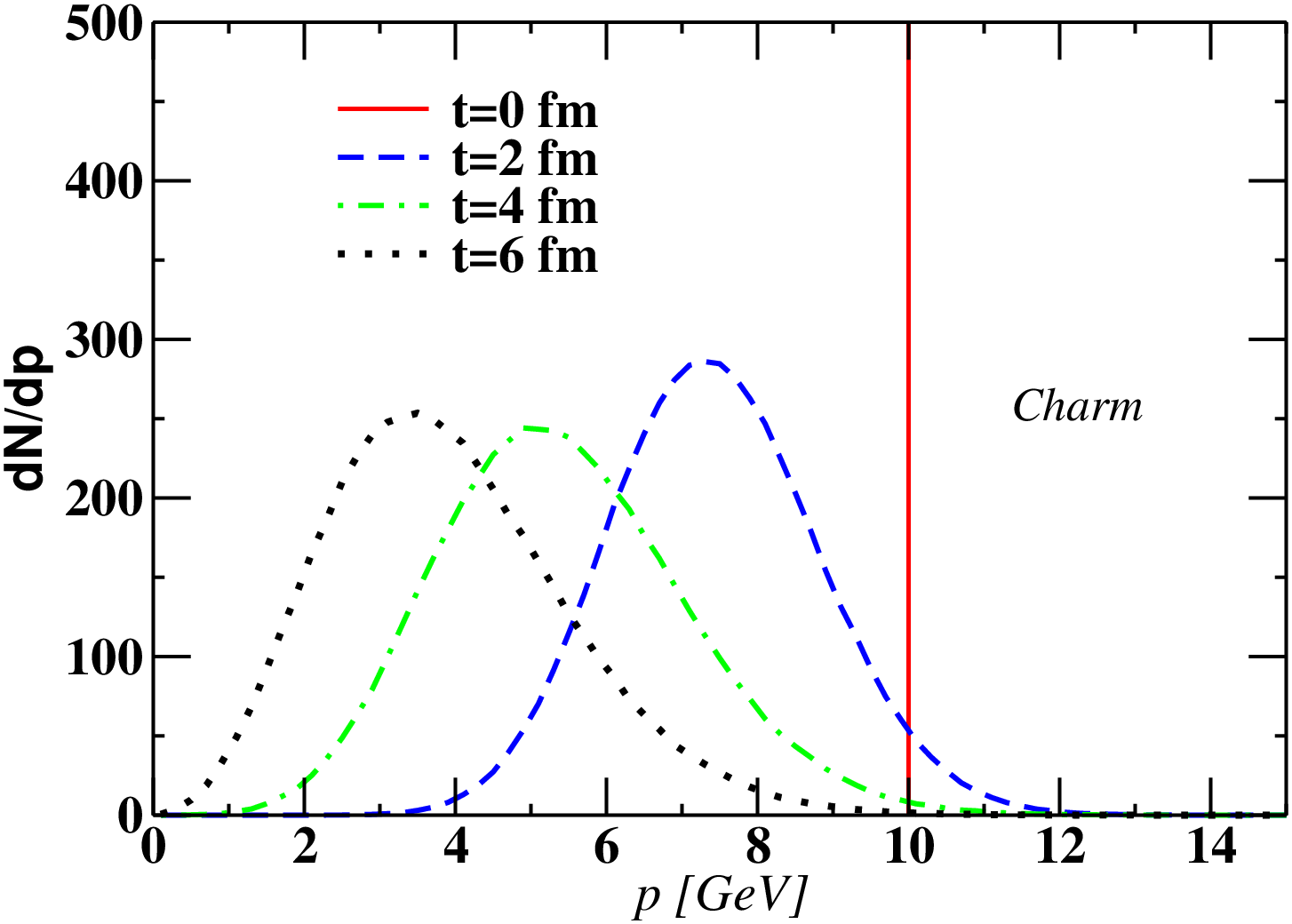}\hspace{2pc}
\caption{Evolution of charm quark momentum distribution within Langevin dynamical  
considering the initial momentum distribution of the charm quarks as a delta 
distribution at p=10 GeV.}
\label{fig9}
\end{center}
\end{figure}

\begin{figure}[ht]
\begin{center}
\includegraphics[width=17pc,clip=true]{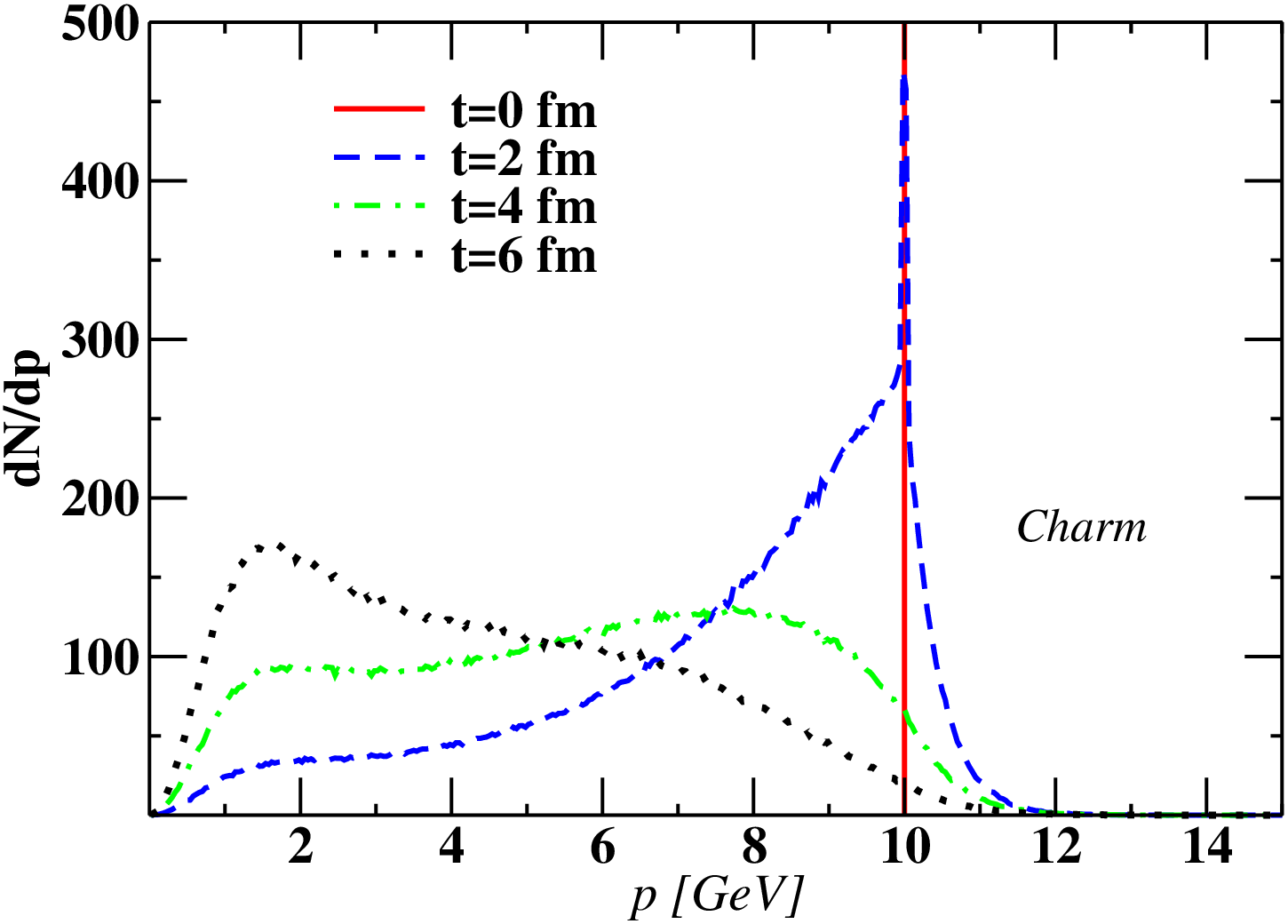}\hspace{2pc}
\caption{Evolution of charm quark momentum distribution within Boltzmann equation  
considering the initial momentum distribution of the charm quark as a delta 
distribution at p=10 GeV.}
\label{fig10}
\end{center}
\end{figure}

\subsection{Momentum spread of heavy quarks}

To further investigate the differences between the heavy quark dynamics implied by 
a Langevin and a Boltzmann approach, we study 
the heavy quark momentum evolution considering the initial charm and bottom quark 
distribution as a delta distribution at $p=10$ GeV for the case with $m_D=0.83$ GeV.  
The momentum evolution of the charm  quarks is displayed in 
Fig.~\ref{fig9} within the Langevin dynamics. 
It is observed that the distributions  are Gaussian as expected
by construction. As known the Langevin dynamics consists of a shift of the average momenta
with a fluctuation around such a value that includes also the possibility to gain energy for the HQ
as we see from the tail of the momentum distribution that overshoots the initial momentum $p=10\,\rm GeV$
at $t=2\,\rm fm/c$, blue dashed line in Fig.\ref{fig9}.

\begin{figure}[ht]
\begin{center}
\includegraphics[width=17pc,clip=true]{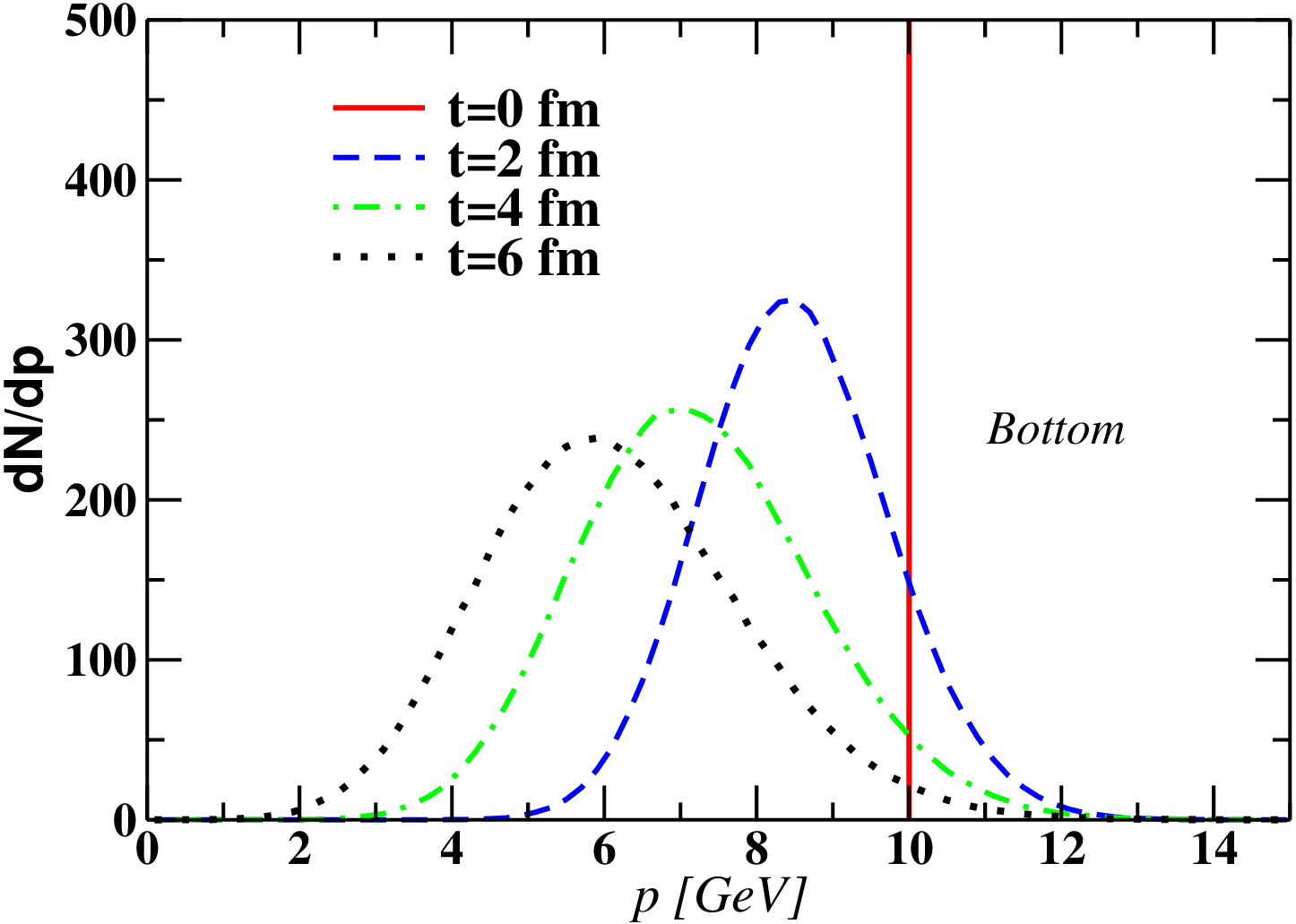}\hspace{2pc}
\caption{Evolution of bottom quark momentum distribution within Langevin dynamical  
considering the initial momentum distribution of the bottom quark as a delta 
distribution at p=10 GeV.}
\label{fig11}
\end{center}
\end{figure}
\begin{figure}[ht]
\begin{center}
\includegraphics[width=17pc,clip=true]{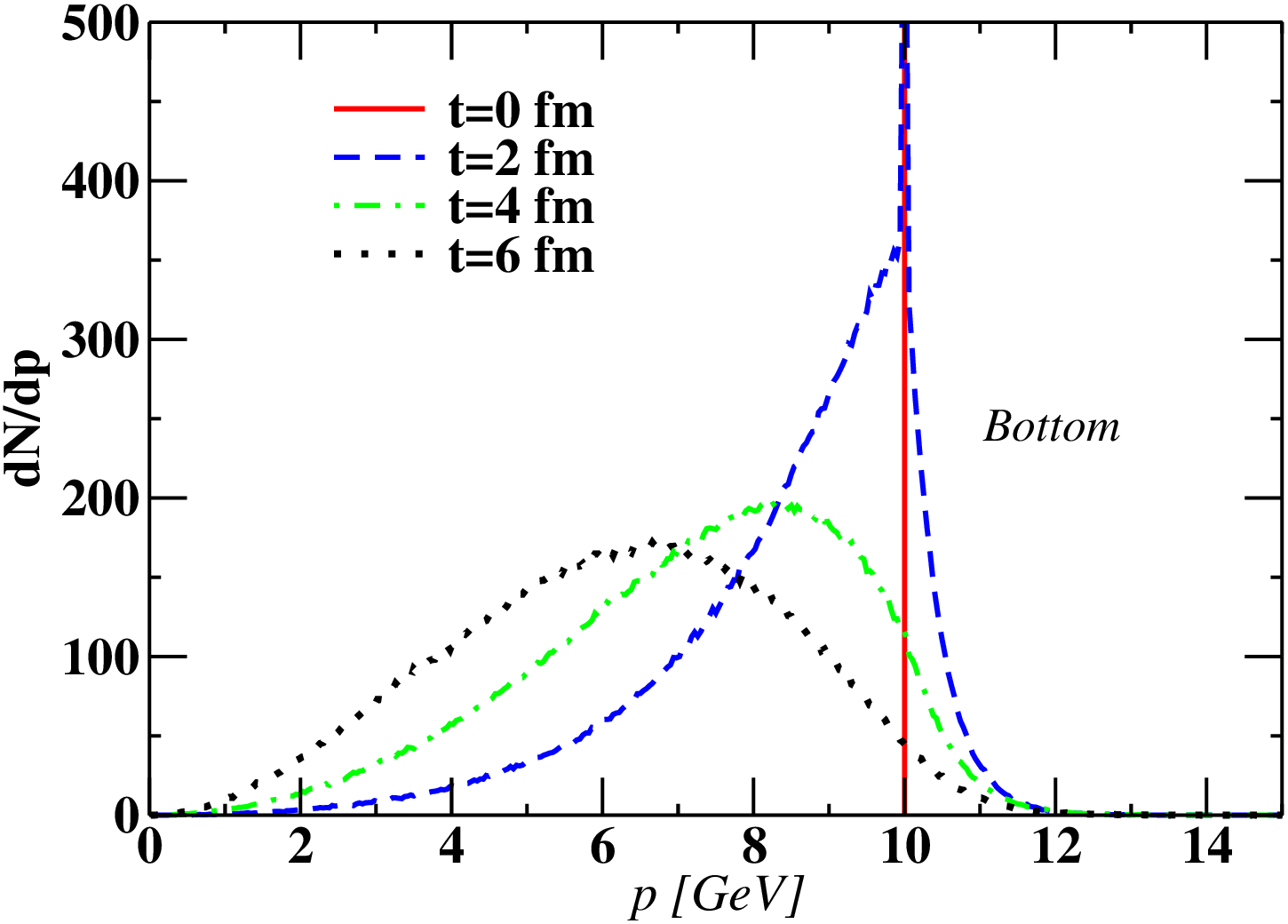}\hspace{2pc}
\caption{Evolution of bottom quark momentum distribution within Boltzmann equation  
considering the initial momentum distribution of the bottom quark as a delta 
distribution at p=10 GeV.}
\label{fig12}
\end{center}
\end{figure}
\begin{figure}[ht]
\begin{center}
\includegraphics[width=17pc,clip=true]{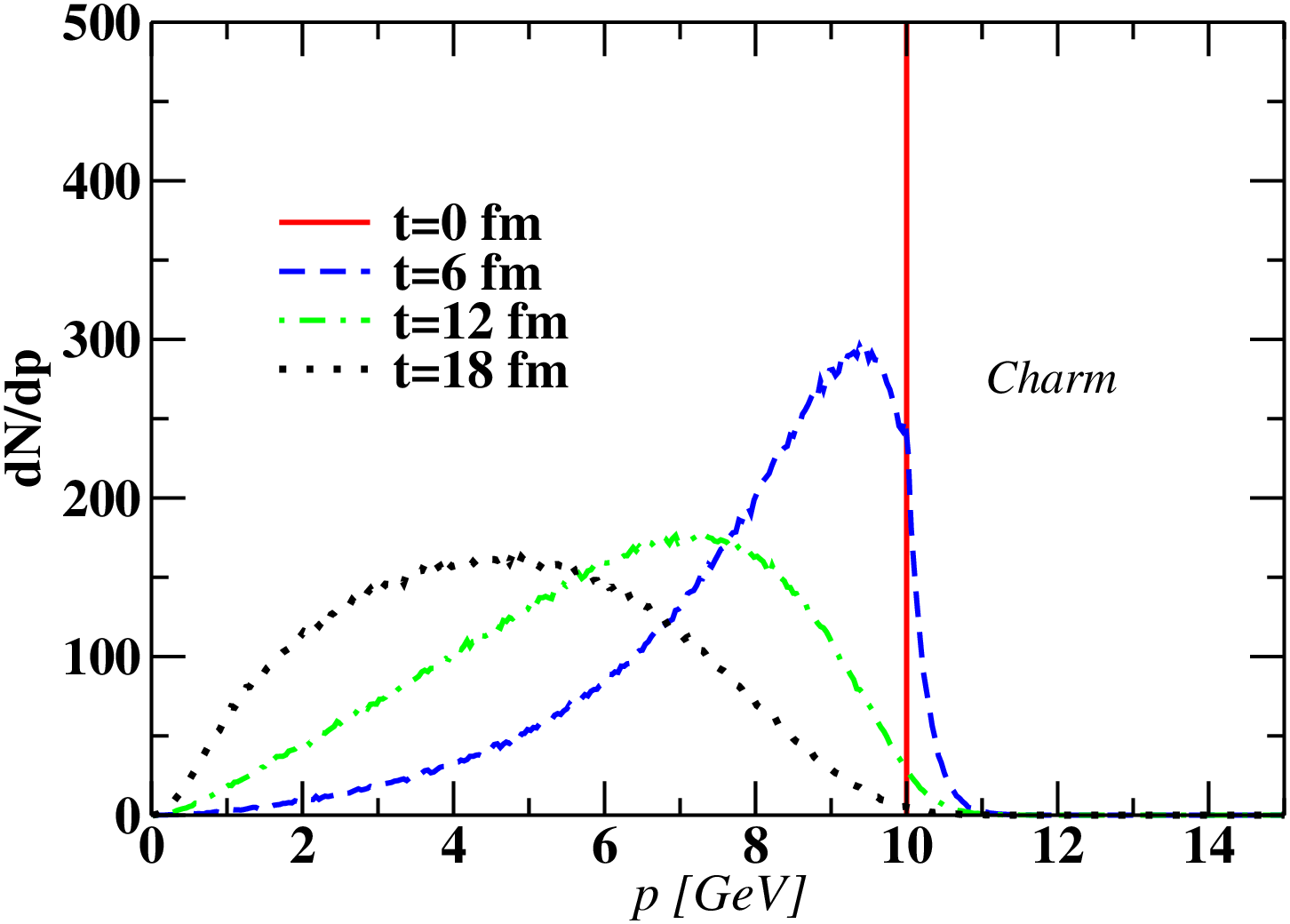}\hspace{2pc}
\caption{Evolution of charm quark momentum distribution using the Boltzmann approach  
considering the initial momentum distribution of the charm quark as a delta 
distribution at p=10 GeV propagating in a bulk at $T=200 MeV$.}
\label{dndp_deltat200}
\end{center}
\end{figure}
In Fig.~\ref{fig10} we present the momentum distribution for charm quark within the Boltzmann 
equation. In this case the evolution of the charm quarks momentum  does not have a Gaussian shape
and already at $t=2\,\rm fm/c$ has a very different spread in momentum with a larger contribution from processes
where the charm quark can gain energy and a long tail at low momenta corresponding to some probability 
to lose a quite large amount of energy
and in general a global shape that is not a at all of Gaussian form. 
This essentially indicates that for a particle with $M \sim \langle p\rangle \sim 3T$ as it is for the charm 
quark at a temperature
$T= 0.4\,\rm GeV$, the evolution is not of Brownian type.
For the bottom quarks, shown in Fig.~\ref{fig11} and Fig.~\ref{fig12} , the momentum evolution gives 
a much better agreement between the Boltzmann
and the Langevin evolution because $M_{b}/T \simeq 10$. To further 
support this argument we have also studied the evolution of charm in a medium at  
$T=200 MeV$ as shown in Fig.~\ref{dndp_deltat200}, where $M_{c}/T \simeq 6$,
 and in this case we have  also found for the charm quark a momentum distribution similar 
to the bottom one shown in Fig.~\ref{fig12}.
We notice that in Fig.~\ref{dndp_deltat200} we have plotted the momentum 
distribution at larger time steps $t_{i}$ with respect to the figures at $400$ MeV. 
This is because the 
drag coefficient $A$ at $200$ MeV is about a factor three smaller than the case at $400$ MeV. 
Therefore, we have chosen to plot 
the distribution at time steps such that  $t_{i}A$ is almost the same as in the 
previous Fig.~\ref{fig9} and \ref{fig10}.\\

Finally, we notice that even in the bottom case at $T= 0.4\,\rm GeV$ ($M_{b}/T \simeq 10$)  the 
evolution of the global spectra are practically
identical between the Langevin and Boltzmann dynamics, see Fig.\ref{fig8},
the detail of the energy loss of a single bottom quark remains still significantly different.
The momentum distribution is reminiscent of a Gaussian distribution showing clearly a peak around
the average momentum, but still it has an asymmetric distribution with a long tail towards lower
momenta.

It would be interesting to study observables that are sensitive to such details of the HQ dynamics.
A first candidate could be 
the $D\bar{D}$ and/or $B\bar{B}$ correlation~\cite{zhuang} that should be quite different in a Langevin 
dynamics with respect to the Boltzmann one since the momentum evolution of a single quark is so different, in particular
for charm quarks.
For the charm quark the very different change in momenta could determine also a quite different
dynamics for the suppression of charmonium in the medium.

\subsection{Time Evolution of the Nuclear Modification factor $R_{AA}$ }

One of the key observable, investigated at RHIC and LHC energies, is the 
depletion of high $p_T$ particles (D and B mesons or single $e^\pm$)
produced in Nucleus-Nucleus  collisions with respect to those produced in proton-
proton collisions expressed through the nuclear suppression factor  $R_{AA}$. 
Therefore, we look at the evolution of the spectra in terms of the
$R_{AA}(p)$ for charm quarks evolving according to the LV and BM transport equations.
We calculate the nuclear suppression factor, $R_{AA}$,  using our initial $t=0$ 
and as final the $t=t_f$ charm quark distribution as $R_{AA}(p)=\frac{f(p,t_f)}{f(p,t_0)}$.

\begin{figure}[ht]
\begin{center}
\includegraphics[width=17pc,clip=true]{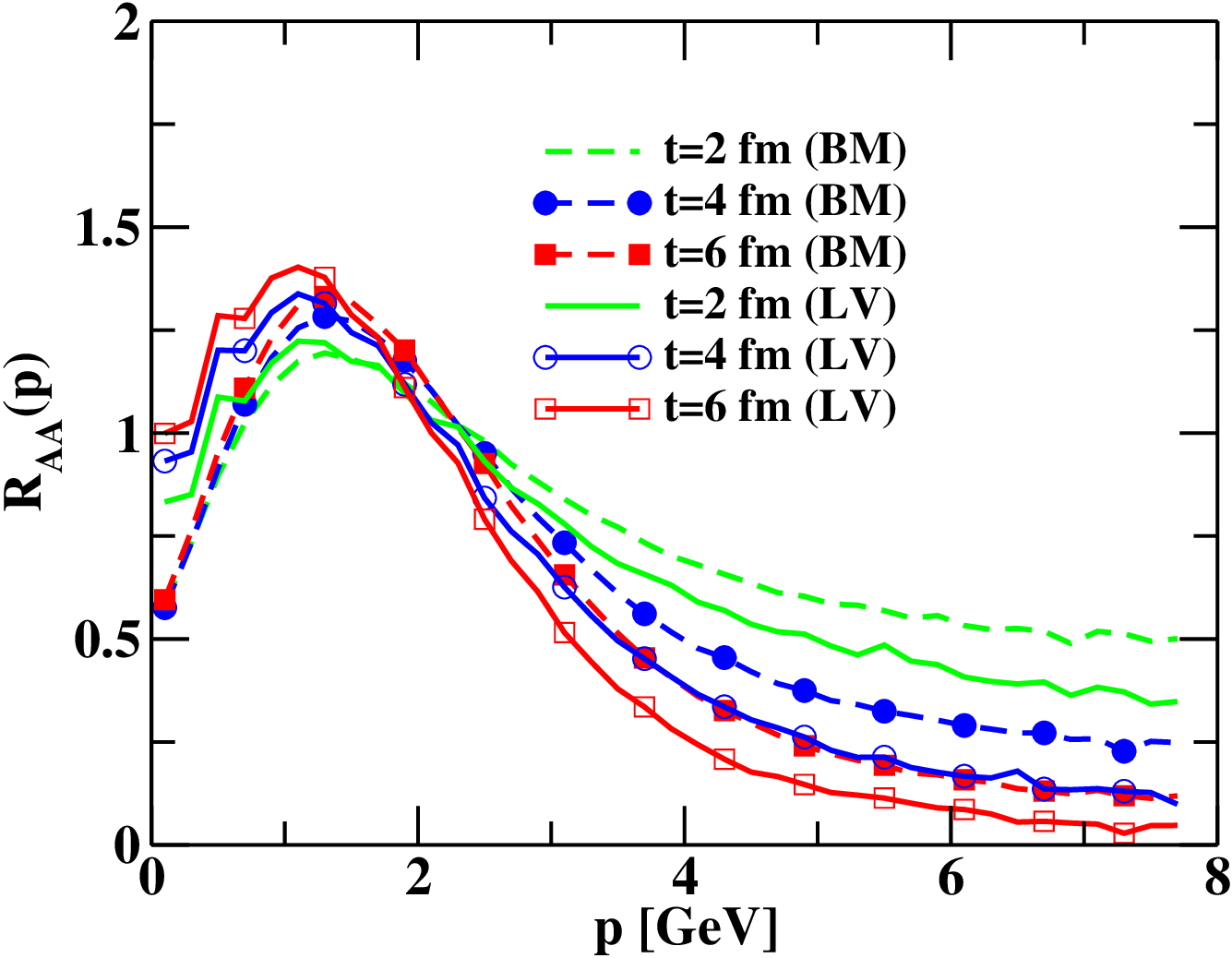}\hspace{2pc}
\caption{The nuclear suppression factor, $R_{AA}$ as a function of 
momentum from the Langevin (LV) equation and Boltzmann (BM) equation for charm quark in a box at T=0.4 GeV 
and $m_D=0.83$ GeV.}
\label{fig13}
\end{center}
\end{figure}

\begin{figure}[ht]
\begin{center}
\includegraphics[width=17pc,clip=true]{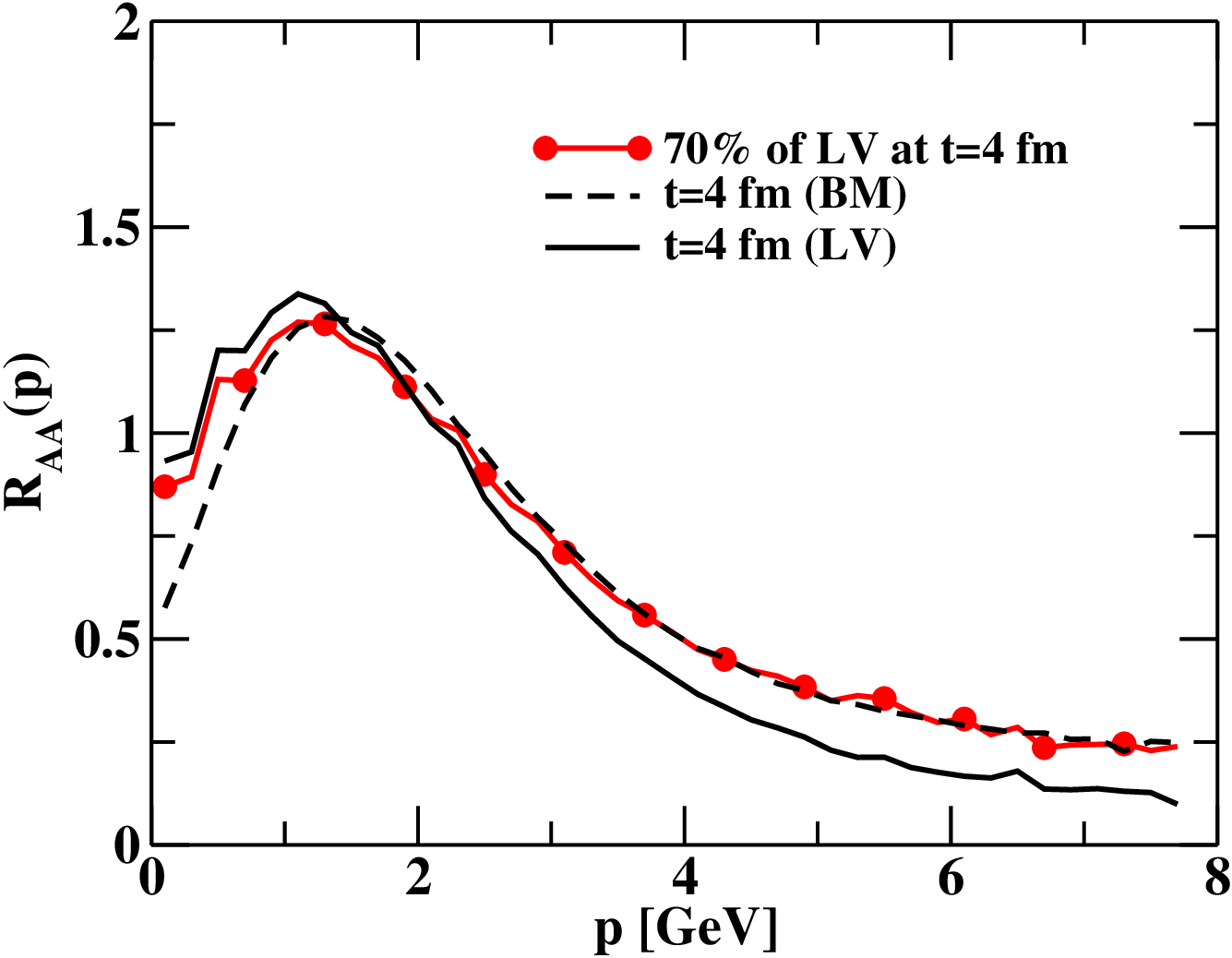}\hspace{2pc}
\caption{The nuclear suppression factor, $R_{AA}$ as a function of 
momentum from the Langevin (LV) equation and Boltzmann (BM) equation for charm quark in a box at T=0.4 GeV 
and $m_D=0.83$ GeV.}
\label{fig14}
\end{center}
\end{figure}


\begin{figure}[ht]
\begin{center}
\includegraphics[width=17pc,clip=true]{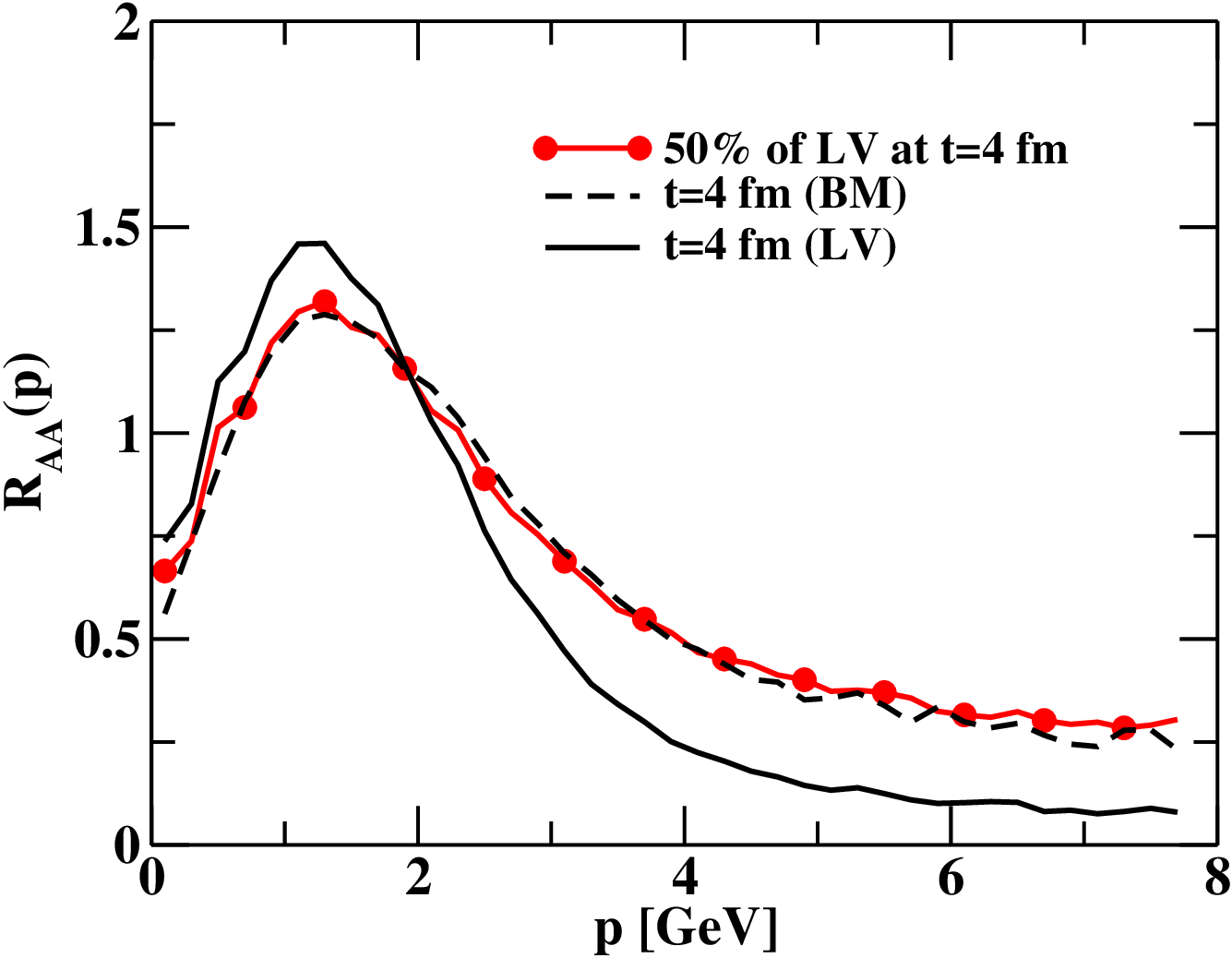}\hspace{2pc}
\caption{The nuclear suppression factor, $R_{AA}$ as a function of 
momentum from the Langevin (LV) equation and Boltzmann (BM) equation for charm quark in a box at T=0.4 GeV 
and $m_D=1.6$ GeV.}
\label{fig15}
\end{center}
\end{figure}

The nuclear suppression factor, $R_{AA}$, has been displayed in Fig~\ref{fig13} as 
a function of momentum from both Langevin and Boltzmann side at different time for $m_D=0.83$ GeV. 
From Fig~\ref{fig13} it is observed that the time evolution of the nuclear suppression factor differs substantially 
from Langevin to Boltzmann at a given time. 
Similar trends are seen also at $m_D=0.4$ GeV but with a much smaller deviation of the order of $15\%$
while at $m_D=1.6$ GeV the deviation in the time evolution are even larger, we will discuss
them more quantitatively before closing this Section.

The Diffusion coefficient has a significant importance for the phenomenological 
study. Hence, it is more meaningful from a phenomenological point of view to evaluate how much 
we need to change the diffusion
coefficient/interaction from Langevin side to reproduce the same nuclear suppression factor as for the Boltzmann 
equation. 
We find that (Fig.~\ref{fig14}) a reduction of the diffusion coefficients is needed for the Langevin equation 
by $30\%$  (or we need only $70\%$ in the LV ) to get a similar nuclear suppression 
factor as of the Boltzmann equation at time $t=4$ fm/c 
which is the typical lifetime of the system produced at RHIC. 
We do not display the experimental data because we are simply studying the evolution of $R_{AA}$
in a box at fixed $T$. Still one gets an idea of the differences that may be found with a realistic
expanding QGP background .
It is interesting to note that even if we put the emphasis on the different dynamical evolution between BM 
and LV, at the end one can obtain an 
identical (almost) $R_{AA}(p)$ in all the $p$ range just by reducing the interaction of about $30\%$.
It may be mentioned here that in the present calculation of $R_{AA}$, both the 
drag and diffusion coefficients are taken from pQCD calculation to capture the actual 
dynamical evolution. It is found that if we use the diffusion coefficient from pQCD and drag 
coefficient in accordance with Einstein relation, the $R_{AA}$  differ about $25\%$ at $p\sim 6-7$ GeV 
and bit less at low momentum 
for $t=4$ fm/c with respect to the case when both the drag and diffusion coefficients are taken 
from pQCD, as we do in this paper. 

In Fig~\ref{fig15} we plotted $R_{AA}$ as a function of momentum from for $m_D=1.6$ GeV. 
For this case the Langevin results differ drastically from Boltzmann as the momentum 
transfer is very large, but still one can  compensate the difference between Langevin and Boltzmann,
having exactly the same $R_{AA}(p)$ by reducing by about a $50\%$ of the diffusion coefficient for the LV evolution.
However, as mentioned in the previous subsection for $m_D=1.6$ GeV there can be significant
dependency whether both A and D coefficients are taken from the pQCD calculations or the fluctuation-dissipation theorem
is implemented. We have checked that this can reduce or enhance the differences between the LV
and BM dynamics. However, our main aim here is simply to show that even if the underlying
dynamics of the charm quark can be quite different between the LV and the BM transport approaches
at the level of the $R_{AA}(p)$, one can anyway mimic the same result mocking
the differences in the dynamical evolution by modifying the interaction by an amount that
can go from about a $10\%$ up to about a $50\%$ depending on the angular dependence of the
scatterings that entails the strength of the transferred momentum.
We have not shown results for the bottom case because, as discussed in the first part of this section
we do not observe significant differences in the time evolution of the spectra between
the LV and BM descriptions.

\section{Comparison with the Experimental observables}
In order to study the impact of the results presented in the previous sections on the  phenomenology of heavy-ion 
collisions, we present here a first comparison 
between the results obtained within the Boltzmann and the Langevin approach with the experimental data. 
We have performed simulation of  $Au+Au$ collisions at $\sqrt{s}= 200$ AGeV
for the minimum bias using a 3+1D transport approach~\cite{greco_cascade,Ruggieri:2013bda,Scardina:2012hy}.
The initial conditions for the bulk in the $r$-space are given by the standard Glauber condition, while in the $p$-space
we use a Boltzmann-Juttner distribution function up to a transverse momentum $p_T=2$ GeV and
at larger momenta mini-jet distributions as calculated by pQCD at NLO order \cite{Greco:2003xt}. 
The initial maximum temperature at the center of the fireball is $T_0=340$ MeV and
the initial time for the simulations is $\tau_0=0.6$ fm/c 
(corresponding to the $\tau_0 \cdot T_0 \sim 1$ criteria). 
The distributions in p-space for the HQ are the same as described in the previous section and in r-space
they are distributed according to $N_{coll}$.
Of course the same bulk that we get within the Boltzmann equation is used
for both the Langevin and Boltzmann approaches. 

\begin{figure}[ht]
\begin{center}
\includegraphics[width=17pc,clip=true]{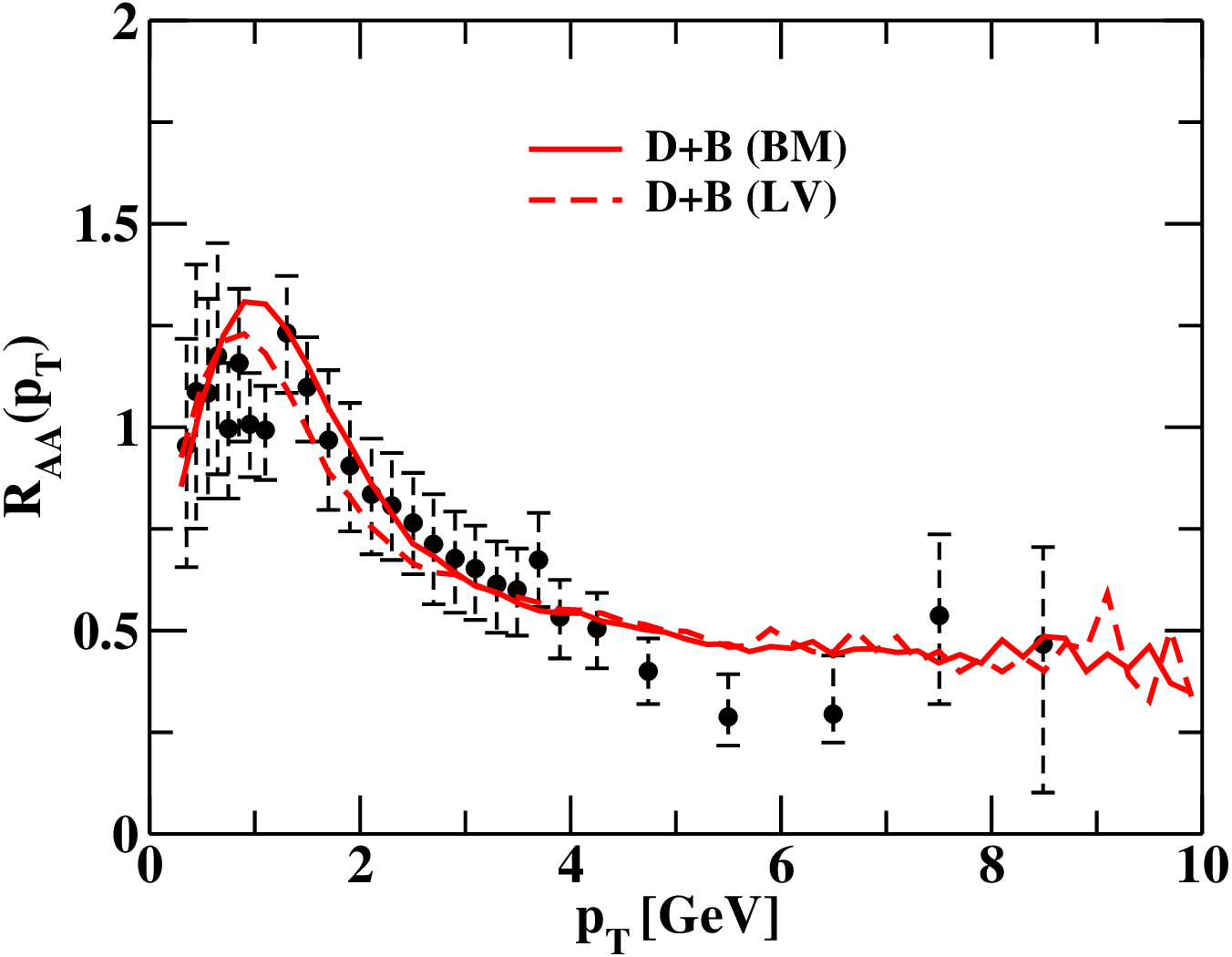}\hspace{2pc}
\caption{Comparison of the nuclear suppression factor, $R_{AA}$, as a function 
of $p_T$ obtained within the Boltzmann (BM) and Langevin (LV) evolution 
with the experimental data obtained at RHIC energy.}
\label{fig16}
\end{center}
\end{figure}

\begin{figure}[ht]
\begin{center}
\includegraphics[width=17pc,clip=true]{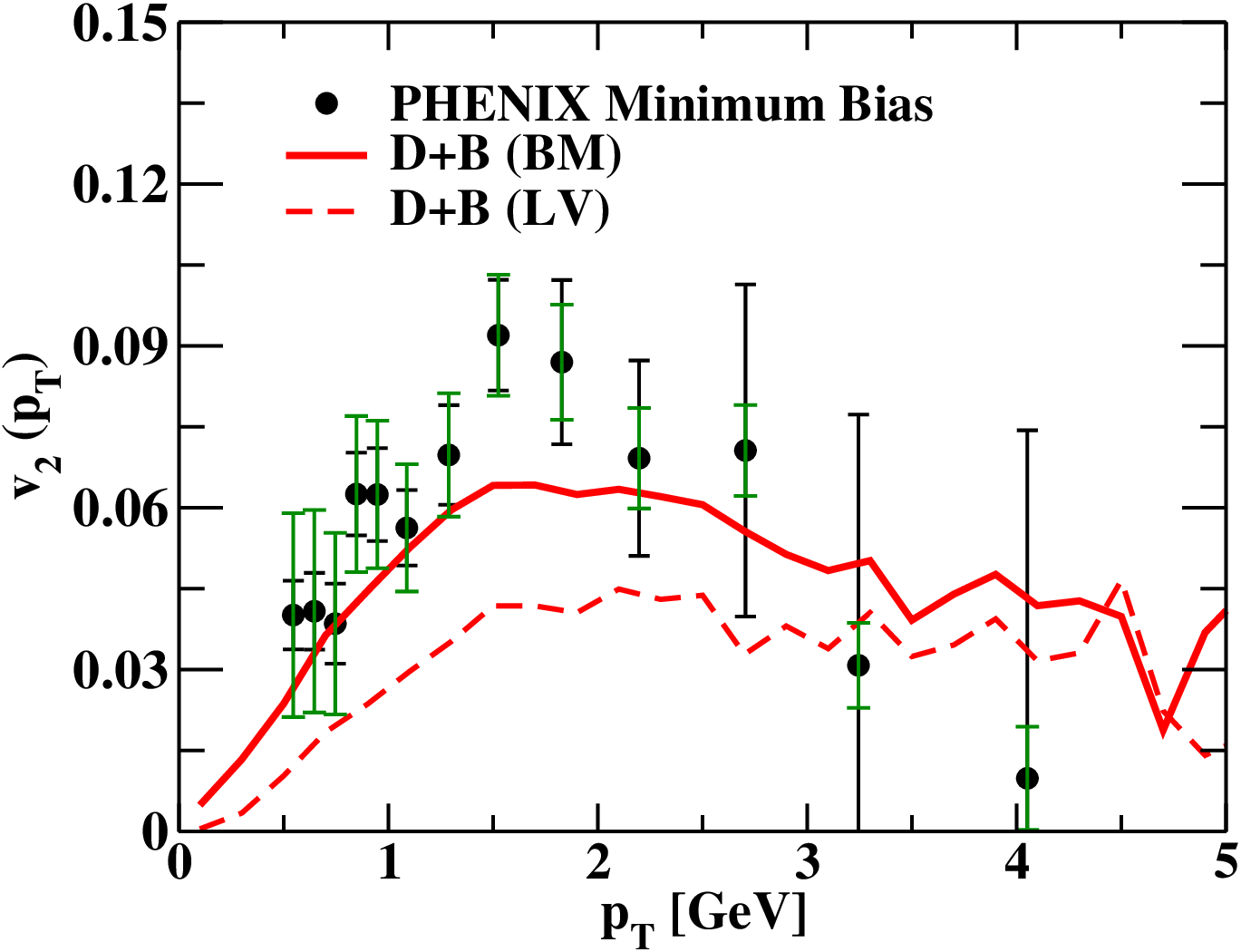}\hspace{2pc}
\caption{Comparison of the elliptic flow , $v_2$, as a function 
of $p_T$ obtained within the Boltzmann (BM) and Langevin (LV) evolution 
with the experimental data obtained at RHIC energy. Both Statistical and systematic errors 
of the experimental data has been taken into accounts. }
\label{fig17}
\end{center}
\end{figure}
 
We convolute the solution with the fragmentation functions 
of the heavy quarks at the transition temperature $T_c$ to 
obtain the momentum distribution of the heavy mesons (B and D).
The Peterson function has been used for heavy quark fragmentation:
\be
f(z) \propto 
\frac{1}{\lbrack z \lbrack 1- \frac{1}{z}- \frac{\epsilon_c}{1-z} \rbrack^2 \rbrack}
\ee
for charm quark $\epsilon_c=0.04$. For bottom quark $\epsilon_c=0.005$.


We calculate the nuclear suppression factor, $R_{AA}$,  using our initial $t=0$ 
and final $t=t_f$ heavy meson (D or B) distribution as $R_{AA}(p)=\frac{f(p,t_f)}{f(p,t_0)}$.
The anisotropic momentum distribution induced by 
the spatial anisotropy of the bulk can be calculated by means of 
the quantity $v_2$:

\be
 v_2=\left\langle  \frac{p_x^2 -p_y^2}{p_T^2}\right\rangle \ , \qquad \qquad
\ee
\label{eq5}

measuring the momentum space anisotropy. 

As we mentioned earlier, it is a challenge for all the models to describe 
the $R_{AA}$ and $v_2$ simultaneously for the same set of inputs.
In Fig.~\ref{fig16} we have shown the $R_{AA}$ as a function of 
$p_T$. In the present study, we try to reproduce the same $R_{AA}$ (almost) from both 
the LV and BM side (as we did in Fig.~\ref{fig14} and Fig.~\ref{fig15} ) and  
study the corresponding $v_2$. To obtain a very similar  $R_{AA}$ in LV and BM for the 
expanding medium, one needs
to reduce the interaction from the LV side similarly to the static case.

In Fig.~\ref{fig17} we have plotted $v_2$ as a function of $p_T$ calculated from both the LV and BM dynamics. 
Our main conclusion is that even if the $R_{AA}(p_T)$ is very similar in  both cases, the full BM 
dynamics generates a sizable larger
$v_2(p_T)$ (about $50\%$ ) toward a better agreement with the experimental data at RHIC ~\cite{phee} energy. 
The present calculation 
is performed at $m_D$=1.6 GeV (isotropic cross section) which allows to simulate the case of nearly isotropic
cross section estimating roughly the maximum effect that the BM dynamics can have
with respect to the LV dynamics for experimental observables. In BM case, we are getting a larger $v_2$, most likely 
due to the large spreading of momentum, see Fig.~\ref{fig10}, implying  that the charm quark 
mixes up with the bulk participating 
more effectively to the dynamics of the expanding medium.  
The effect on $v_2$ for the same $R_{AA}(p_T)$ is still 
significant, about $25\%-30\%$, for $m_D \sim 0.8 \rm\, GeV$ 
(similarly for $m_D=g(T) \, T$), while it becomes negligible for $m_D \sim 0.4 \rm\, GeV$ (or more generally
for $m_D \leq T$). 
We notice that with the isotropic cross section one may nearly reproduce the $R_{AA}$ and $v_2$ 
simultaneously within the BM approach.

\section{Summary and conclusions}

We present a study 
of the implications of the approximations involved in the Langevin equation by mean of a direct 
comparison with the full collisional integral (no small momentum transferred approximation)  
within the framework of the Boltzmann transport equation. We consider a box where the bulk is in equilibrium 
at T=0.2-0.4 GeV. For the realistic initial momentum distribution, we found that the 
Langevin approach is a very good approximation for bottom quark whereas for charm quark 
Langevin approach can deviate significantly from the full Boltzmann transport 
equation depending on different values of $m_D$.
The difference for $m_D=0.83$ GeV is about a $40-50\%$ at intermediate momentum for a time evolution of 5-6 fm/c typical 
of the QGP created in heavy-ion collisions . The deviation goes down to
about a $10-15\%$ for the same range of momentum and time for the case $m_D=0.4$ GeV. 
For $m_D=1.6$ GeV the difference is about a $70\%$ at intermediate momentum,
but the exact amount depends on the way the fluctuations-dissipation theorem is implemented.
Hence, the effect for charm quarks can
be significantly larger or smaller depending on the values of Debye screening mass 
and/or on the angular dependence of the collisions.
However we  notice that one can get a very similar suppression factor from both the approaches 
just reducing the diffusion coefficient of the Langevin approach by around $30\%$ for $m_D=0.83$ GeV. 
For $m_D=1.6$ GeV such reduction should be about a $50\%$.

The present uncertainties on the charm transport coefficients is indeed even larger. 
If such an ambiguity on the value of the drag or diffusion coefficient is discarded looking only
at $R_{AA}(p_T)$ one can conclude the Langevin dynamics supplies a sufficiently good approximation
of the Boltzmann collision integral, even if we warn that the detail of the single quark energy loss spread could be
quite different in the two cases.
In fact, we have also performed the comparison for the case considering the initial distribution 
as a delta peaked at $p=10$ GeV. We found that both the charm and bottom quarks are Gaussian 
like distribution within the Langevin dynamics while the distribution obtained, for 
charm quarks, within the Boltzmann equation is a quite different one revealing the 
fact that Langevin dynamics of the charm quark may not work in a hot QCD medium. 
For the bottom quarks the approximation looks reasonable.

More specifically, we have found that the underlying dynamics of the single heavy quarks are indeed
quite different between the Boltzmann and the Langevin approach. In practice 
the dynamical evolution
of a single charm does not appear to be simply a shift in momentum with a Gaussian fluctuations around
it. Indeed one can say that this can be expected considering that for charm quarks $M_c\sim \langle p \rangle$
for a typical temperature at RHIC and LHC energies, which makes the assumption of Brownian motion questionable. 
On the contrary, for bottom quarks we do find that the dynamical evolution is approximately that
of Brownian motion even if the fluctuation around the average momentum still appear to deviate from
a Gaussian shape and are more reminiscent of the binomial distribution.
These results can have significant effects on the heavy ion phenomenology 
at RHIC and LHC energies.

Finally we presented 
a first comparison between LV and BM approach for a realistic fireball evolution 
as created in ultra-relativistic heavy-ion collisions at RHIC energy. We have found that 
with the isotropic cross section the BM approach 
gets close to reproducing the $R_{AA}$ and $v_2$ 
simultaneously while the LV dynamics generate a smaller elliptic flow in correspondence to the same $R_{AA}$.

\vspace{2mm}
\section*{Acknowledgments}
We acknowledge the support by the ERC StG under the QGPDyn
Grant n. 259684. We acknowledge to T. Bhattacharyya for reading the manuscript.   
\vspace{3mm}

\section{Appendix}
In this appendix we quote the invariant amplitude for the elastic processes
used for evaluating the drag and diffusion coefficients of charm and bottom quarks 
considering the bulk consists of only gluons. 

The invariant amplitude,
$|\mathcal{M}|^2_{gHQ\rightarrow gHQ}$
 for the process $gHQ\rightarrow gHQ$ is given by:
 
\begin{eqnarray}
|{\cal M}_{gHQ \rightarrow gHQ}|^2={\pi^2 \alpha_s^2} 
\bigg\lbrack 
\frac{32(s-M^2)(M^2-u)}{(t-m_D^2)^2}&& \nonumber  \\ 
+\frac{64}{9}\frac{(s-M^2)(M^2-u)+2M^2(s+M^2)}{(s-M^2)^2} \nonumber  \\ 
+\frac{64}{9}\frac{(s-M^2)(M^2-u)+2M^2(M^2+u)}{(M^2-u)^2}\nonumber\\
+ \frac{16}{9}\frac{M^2(4M^2-t)}{(s-M^2)(M^2-4)} \nonumber  \\   
+16\frac{(s-M^2)(M^2-u)+M^2(s-u)}{(t-m_D^2)(s-M^2)}\nonumber\\
-16\frac{(s-M^2)(M^2-u)-M^2(s-u)}{(t-m_D^2)(M^2-u)} 
\bigg\rbrack 
\end{eqnarray}
where s,t and u are the  Mandelstam variables and $M$ is the mass of the heavy quark.
Integrating the invariant amplitude over the variable t, between the integration
limits 
\be
t_{min}=-\frac{(s-M^{2})^2}{s} ; t_{max}=0
\ee
we get the total cross section 
\be
\sigma_{gHQ \rightarrow gHQ}=\frac{1}{16\pi(s-M^{2})^2}\int^{t_{max}}_{t_{min}}dt|{\cal M}_{gHQ \rightarrow gHQ}|^2
\ee 
that we need  to evaluate the collision integral in eq. \ref{B_E}.

\end{document}